\RequirePackage[T1]{fontenc}
\documentclass[12pt]{article}

\usepackage[height=8.85in,width=6.45in]{geometry}

\usepackage[utf8]{inputenc}
\usepackage{amsmath}
\usepackage{amssymb}
\usepackage{mathtools}
\usepackage{physics}
\numberwithin{equation}{section}
\usepackage{slashed}
\usepackage{braket}
\usepackage[svgnames,psnames]{xcolor}
\usepackage[colorlinks,citecolor=DarkGreen,linkcolor=FireBrick,linktocpage]{hyperref}

\usepackage{varioref}       
\usepackage{cleveref}  

\crefname{table}{table}{tables}
\Crefname{table}{Table}{Tables}
\crefname{figure}{figure}{figures}
\Crefname{figure}{Figure}{Figures}

\usepackage{cite}
\usepackage{graphicx}
\usepackage{tikz}
\usepackage{tikz-cd}
\usepackage{times}
\usepackage{courier}
\usepackage{bm}
\usepackage{subfig}
\usepackage{ascmac}
\usepackage{tikzsymbols}
\usepackage{tensor}

\usepackage{mdframed}

\renewenvironment{figure}[1][]{
  \begin{originalfigure}[#1]
    \begin{mdframed}[linecolor=black!0,backgroundcolor=black!1]
}{
    \end{mdframed}
  \end{originalfigure}
}
\renewenvironment{table}[1][]{
  \begin{originaltable}[#1]
    \begin{mdframed}[linecolor=black!0,backgroundcolor=black!1]
}{
    \end{mdframed}
  \end{originaltable}
}

\usepackage{colortbl}
\definecolor{lightyellow}{rgb}{1.0, 0.95, 0.7}
\definecolor{lightblue}{rgb}{0.7, 0.9, 1.0}
\definecolor{lightpink}{rgb}{1.0, 0.85, 0.95}
\definecolor{lightgreen}{rgb}{0.7, 1.0, 0.4}

\def\sp{\text{sp}}

\def\boo{0.0}

\def\xlattice#1#2#3{
\begin{tikzpicture}[scale=.5]
\filldraw[color=black!5!white](-.5,-.5) rectangle (1.5,1.5);
\draw[->] (-1,0) -- (2,0);
\draw[->] (0,-1) -- (0,2);
\foreach \x in {0,1} {
	\foreach \y in {0,1}{
		\pgfmathsetmacro\a{mod(#1 * \x - #2 * \y,2)}
		\ifx\a\boo
			\filldraw[color=#3] (\x,\y) circle (.5em);
		\else
			\filldraw[fill=white,draw=gray] (\x,\y) circle (.5em);
		\fi
	}
}
\end{tikzpicture}
}

\newenvironment{eqaed}
    {\begin{equation}
    \begin{aligned}
    }
    { 
    \end{aligned}
    \end{equation}
    \ignorespacesafterend
    }

\begin{document}

\begin{titlepage}

\begin{flushright}
LMU-ASC 07/24\\
IFT-UAM/CSIC-24-70
\end{flushright}

\vskip 3cm

\begin{center}

{\Large \bfseries Species scale, worldsheet CFTs and emergent geometry}

\vskip 1cm
Christian Aoufia$^1$, Ivano Basile$^2$ and Giorgio Leone$^3$
\vskip 1cm

\begin{tabular}{ll}
$^1$ & \emph{Instituto de F\'{i}sica Te\'{o}rica IFT-UAM/CSIC}\\
& \emph{C/ Nicol\'{a}s Cabrera 13-15, Campus de Cantoblanco, 28049 Madrid, Spain} \\
$^2$ & \emph{Arnold-Sommerfeld Center for Theoretical Physics}\\ & \emph{Ludwig Maximilians Universit\"at M\"unchen}\\ & \emph{Theresienstraße 37, 80333 M\"unchen, Germany} \\
$^3$ & {\em Dipartimento di Fisica, Universit\`a di Torino and INFN Sezione di Torino} \\
& {\em Via Pietro Giuria 1, 10125 Torino, Italy}
\end{tabular}

\end{center}

\begin{abstract}
    We study infinite-distance limits in the moduli space of perturbative string vacua. The remarkable interplay of string dualities seems to determine a highly non-trivial dichotomy, summarized by the emergent string conjecture, by which in some duality frame either internal dimensions decompactify or a unique critical string becomes tensionless. We investigate whether this pattern persists in potentially non-geometric settings, showing that (a proxy for) the cutoff of the gravitational effective field theory in perturbative type II vacua extracted from a graviton scattering amplitude vanishes if and only if a light tower of states appears. Moreover, under some technical assumptions on the spectrum of conformal weights, the cutoff scales with the spectral gap of the internal conformal field theory in the same manner as in decompactification or emergent string limits, regardless of supersymmetry or whether the internal sector is geometric. As a byproduct, we elucidate the role of the species scale in (de)compactifications and show compatibility between effective field theory and worldsheet approaches in geometric settings with curvature.
\end{abstract}

\end{titlepage}

\setcounter{tocdepth}{2}
\tableofcontents


\section{Introduction}\label{sec:introduction}

It is common folklore that string theory predicts the existence of extra dimensions of space. However, already at the perturbative level the theory offers us a remarkable conceptual generalization of ordinary geometry in the form of (two-dimensional) conformal field theory (CFT). In particular, string backgrounds suitable for an effective field theory description of low-energy gravitational physics describe a parametrically weakly curved\footnote{More precisely, all fields ought to have curvatures parametrically smaller than the string scale. This includes gauge fields and higher form fields, as well as scalars such as the dilaton.} dynamical spacetime $\mathcal{M}_d$ together with some internal sector. Criticality, arising from Weyl anomaly cancellation in the Polyakov path integral for (super)strings, fixes the total central charges, and thus the internal sector ought to have central charges $(c_L, c_R) = (15-\frac{3}{2}d, 26-d)$ in the heterotic case and $(15-\frac{3}{2}d, 15-\frac{3}{2}d)$ in the RNS-RNS case. Such an internal CFT\footnote{Amongst the various consistency conditions, we assume modular invariance of its torus partition function, unitarity and compactness.} defines such a string background, and thus the spacetime dimension $d\leq 10$ is \emph{upper bounded} by 10 in ``tame'' backgrounds. Including the sector described by eleven-dimensional supergravity pushes the bound to 11, but \emph{a priori} it is not clear that the internal CFT be always geometric, namely (perturbatively connected to) a non-linear sigma model on some compact target space. \\

Given this state of affairs, it is perhaps not surprising that the most well-understood examples of two-dimensional CFTs are of this type. Narain theories describing toroidal compactifications and Calabi-Yau sigma models are important examples of \emph{bona fide} perturbative string vacua. These theories, or quotients thereof, can exhibit phases that differ qualitatively from sigma models at special loci in their conformal manifolds \cite{ Kawai:1986va, Narain:1986qm, Lerche:1986cx, Antoniadis:1986rn, Dixon:1987yp,  Candelas:1987kf, Antoniadis:1987wp, Gepner:1987qi, Green:1988wa, Green:1988bp, Candelas:1988di, Kazama:1988qp, Vafa:1988uu, Greene:1988ut, Candelas:1989ug, Candelas:1989js, Narain:1990mw, Witten:1993yc, Kachru:1995wm, Angelantonj:1996mw, Blumenhagen:1998tj, Israel:2013wwa, Hull:2017llx, Gkountoumis:2023fym, Baykara:2023plc}. However, it is not clear whether all non-geometric CFTs can be connected to geometric ones in this fashion. While non-perturbatively there seems to be strong evidence to this effect \cite{McNamara:2019rup, McNamaraThesis, McNamara:swamplandia, Debray:2023yrs}, at the perturbative level such a connection is not \emph{a priori} guaranteed to be present, although in some settings it is possible to provide a CFT description of the phenomenon\footnote{For instance, there are known examples of asymmetric orbifold compactifications \cite{Blumenhagen:2000fp, Angelantonj:2000xf} admitting a T-dual description in terms of geometric orbifolds, still within the perturbative regime amenable to a worldsheet analysis.}. Still, whether the geometric landscape (understood as vacua perturbatively connected to non-linear sigma models) covers the entire set of perturbative string vacua\footnote{Even in supersymmetric settings the problem remains open \cite{Gepner:1987qi, Seiberg:1988pf}.}, geometric settings already provide an extremely rich arena to study string theory. To name but a few examples, Calabi-Yau compactifications, with their added bonus of preserving some spacetime supersymmetry, provide considerable computational control via techniques from algebraic geometry as well as mirror symmetry, which allows to investigate some topology changing effects \cite{Witten:1993yc, Aspinwall:1993yb}. Freund-Rubin backgrounds can afford explicit holographic interpretations, and thus a concrete handle on non-perturbative quantum gravity. The moduli spaces of Narain theories exhibit T-duality and pave the way to orbifold, orientifold, and more elaborate non-geometric constructions \cite{Bianchi:1999uq, Dabholkar:2002sy, Hull:2004in, Dabholkar:2005ve, Shelton:2005cf}. More importantly for our purposes, such dualities non-trivially imply that infinite-distance limits in conformal manifolds of geometric CFTs feature a non-trivial behavior: they can be always described as decompactifications of an internal compact space. Together with the limit in which critical strings are weakly coupled, or asymptotically tensionless in Planck units, up to duality these two behaviors seem to exhaust all behaviors in the string landscape. The emergent string conjecture \cite{Lee:2018urn, Lee:2019wij, Lee:2019xtm}, a strong refinement of the swampland distance conjecture \cite{Ooguri:2006in}, stipulates that indeed these are the only two options. The light spectrum of the theory in such limits reflects this via Kaluza-Klein towers and higher-spin excitations of a unique critical string. \\

The deep physical content of the emergent string conjecture rests crucially on the consistent web of string dualities. In particular, as mentioned above, the existence of a description in terms of decompactifications stems from T-duality, whereas the emergence of a \emph{unique} critical string (even when the starting description has no strings in sight!) in equi-dimensional limits is tied to non-perturbative dualities which relate various sectors of string/M-theory. In this paper we ask whether this dichotomy of infinite-distance limits persists beyond geometric settings -- that is, if indeed geometry always emerges to produce a decompactification limit. The worldsheet approach to string (perturbation) theory is a natural setting to investigate such a question, since one can recast the problem in the language of CFT\footnote{For a recent example in this spirit, see the proof of the weak gravity conjecture given in \cite{Heidenreich:2024dmr}.}. \\

By now there is considerable bottom-up support for the weaker swampland distance conjecture, which stems from the factorization of information metrics in the presence of gravity \cite{Stout:2021ubb, Stout:2022phm}. The existence of infinite towers of degrees of freedom in factorization limits, where the gravitational effective field theory (EFT) becomes weakly coupled all the way to its ultraviolet (UV) cutoff $\Lambda_{\text{UV}} \ll M_\text{Pl}$, is also well-supported from a number of perturbative bootstrap arguments, and its consistency in non-geometric settings was explored in \cite{Demulder:2023vlo}. Hence, we take this as a starting point to investigate the refined dichotomy of limits prescribed by the emergent string conjecture. The combined efforts of \cite{Basile:2023blg, Bedroya:2024ubj} appear to build a convincing bottom-up case in this respect, and in particular the behavior found in \cite{Basile:2023blg} does indicate that any tower whose mass gap $m_\text{gap} \ll \Lambda_\text{UV} \ll M_\text{Pl}$ is below the UV cutoff behaves like a Kaluza-Klein tower. On the other hand, higher-spin stringy excitations must be gapped at or above the UV cutoff, as shown by bootstrap investigations (see also \cite{Martucci:2024trp, Bedroya:2024ubj} for other arguments). Here we focus on a top-down perspective instead, first reviewing the more familiar geometric case and then proceeding to the more abstract CFT setting, where we are able to show that the swampland distance conjecture follows from the hierarchy $\Lambda_\text{UV} \ll M_\text{Pl}$, at least up to some genericity assumption, and that the theory always behaves geometrically in a precise sense. In this context the dichotomy captured by the emergent string conjecture reduces to the emergence of geometry at infinite distance in the worldsheet conformal manifold, since the weak string coupling limit takes into account the alternative option. In more detail, we shall demonstrate that $\Lambda_\text{UV}$, extracted from a four-point graviton scattering amplitude, vanishes if and only if a light tower of states emerges. Moreover, at least under the simplifying technical assumption that the conformal weights of the light states vanish at the same rate, $\Lambda_\text{UV}$ scales with the spectral gap as in geometric decompactification limits. In other words, besides the light tower we shall assume that all other states have conformal weights which are bounded below.\\

The contents of this paper are structured as follows. In \cref{sec:species_compactifications} we review the various relevant scales in (possibly curved) geometric compactifications from the EFT and worldsheet perspectives, paving the way to the more abstract CFT approach of \cref{sec:species_worldsheet}. Therein, we present the main results of this paper: in \cref{sec:light_tower} we derive the existence of a light tower of states assuming that the scale suppressing the $R^4$ terms in type II (possibly) non-geometric vacua vanishes in some limit. Then, under some assumptions on light states which we precisely state, in \cref{sec:light_cutoff} we show that this quantity scales with the spectral gap of the internal CFT with the same power-like law in the central charge(s) expected from geometric decompactification limits, regardless of spacetime supersymmetry. This is a non-trivial necessary requirement for the emergent string proposal of \cite{Lee:2018urn, Lee:2019wij, Lee:2019xtm} to hold, and it is tempting to speculate about the universal emergence of internal geometries. In other words, whenever certain assumptions about the moduli space of string vacua hold, it may be the case that extra dimensions arise in the classical sense from (limits of) the more abstract stringy geometry of CFT. \\

This paper contains three appendices. In \cref{app:heat} and \cref{app:tao} we provide some technical details on the heat-kernel computations employed in \cref{sec:species_compactifications}. Finally, in \cref{app:geometric_vacua} we introduce a systematic worldsheet approach to correct EFT distance computations in the presence of curvature. This allows us to describe the standard version of the distance conjecture of \cite{Ooguri:2006in} and the generalized version of \cite{Lust:2019zwm} in a coherent fashion, ensuring the connection EFT and worldsheet approaches. \\

\textbf{Note added:} during the last stages of completion of this work, a paper with significant overlap in conclusions appeared \cite{Ooguri:2024ofs}, although the discussion therein is framed in terms of the AdS$_3$/CFT$_2$ correspondence rather than string perturbation theory. Besides the derivation of (sharp bounds for) exponential behaviors and infinite distances in limits of light conformal dimensions, most relevantly for us \cite{Ooguri:2024ofs} obtained the important result that the limiting theory contains a ``decompactified'' sigma model. Our results in \cref{sec:species_worldsheet} rather concern the approach to the limit, which also brings along a light tower and behaves in a geometric fashion, indicating the emergence of geometry at infinite distance in the worldsheet conformal manifold. Together with the results of \cite{Ooguri:2024ofs}, the case for the swampland distance conjecture and the emergent string conjecture in (potentially) non-geometric perturbative settings is further reinforced.

\section{Heat kernel asymptotics and the species scale}\label{sec:species_compactifications}

In this section we review some facts about geometric compactifications which we will subsequently generalize in the following. Any $d$-dimensional gravitational EFT with a UV completion is provided with a number of interesting physical scales, signaling the onset of physical phenomena such as strongly coupled gravitational effects or the breakdown of the effective description. In a generic theory all these scales coincide parametrically with the ($d$-dimensional) Planck scale $M_\text{Pl}^{(d)}$, but in the presence of many light species in the spectrum, such as Kaluza-Klein modes or weakly coupled string excitations, various interesting scales can arise \cite{Veneziano:2001ah, Dvali_2010, Dvali:2007wp, Dvali:2010vm} and separate, with the Planck scale an upper bound to all of them.

The resulting scales are commonly and collectively dubbed ``species scale'', although they refer to \emph{a priori} distinct scales. For the purposes of disambiguation, the quantity originally defined of \cite{Veneziano:2001ah, Dvali_2010} may be referred to as the ``counting'' species scale, here and in the following denoted $\Lambda_\sp$, and it arises as an upper bound to the scale $\Lambda_\text{QG}$ of strongly coupled quantum gravity. The original one-loop estimate of the graviton propagator (or some other gravitational quantity in the EFT), which extends from weakly coupled species to conformal field theories\footnote{A middle-ground extension to interacting, non-conformal species is discussed in \cite{Basile:2024dqq}. More generally, a tower may involve non-trivial graviton couplings to each level, which can modify the \emph{effective} number of species contributing to scattering or black-hole evaporation \cite{Dvali:2009ks}.}, gives a ``worst-case scenario'' scale at which perturbation theory fails. The resulting self-consistent counting over species yields the defining (parametric) equation
\begin{eqaed}\label{eq:speciesdef}
    \Lambda_\text{sp}^{2-d} = N_\text{sp} \equiv \sum_{m \leq \Lambda_\text{sp}} 1,
\end{eqaed}
in $d$-dimensional Planck units, where $N_{\text{sp}} \equiv N(\Lambda_\sp)$ is the number of degrees of freedom with mass $m$ below such scale. The definition is parametric, as befits a physical scale, and no $\mathcal{O}(1)$ factors matter in its determination. If the mass gap $m_\text{gap}$ of a parametrically large number of species is parametrically smaller than the Planck scale, \cref{eq:speciesdef} results in the hierarchy $m_\text{gap} \ll \Lambda_\sp \ll M_\text{Pl}^{(d)}$. \\

Another relevant scale for any EFT is of course the UV cutoff $\Lambda_\text{UV}$\footnote{More precisely, we shall concern ourselves with the UV cutoff of its gravitational sector, as in \cite{Cribiori:2022nke, Cribiori:2023sch, vandeHeisteeg:2022btw, vandeHeisteeg:2023ubh, Castellano:2023aum, vandeHeisteeg:2023dlw, Bedroya:2024uva, Bedroya:2024ubj}. The ``strict'' EFT with all massive species integrated out is of course cut off at $m_\text{gap}$, where massive species compromise unitarity of the restricted S-matrix and manifest additional higher-dimensional saddles in the semiclassical path integral. Such EFTs may not retain gravitational physics at infinite distance and become rigid \cite{Marchesano:2023thx}.}, which is upper bounded by $\Lambda_\text{QG}$ since quantum gravity is weakly coupled at low energies. In the expected scenario of finitely many fine-tunings \cite{Heckman:2019bzm} of Wilson coefficients, the usual definition of $\Lambda_\text{UV}$ is that of the smallest scale parametrically suppressing all but finitely many higher-derivative operators in the effective action. A sharper definition which circumvents potential issues due to field redefinitions trades higher-derivative operators with the corresponding kinematic forms in the low-energy expansion of scattering amplitudes (or AdS boundary correlators), which are well-defined ``holographic'' observables. Therefore, since the gravitational sector of the EFT is weakly coupled below its UV cutoff\footnote{One way to see this is that the quantum couplings of gravitons are proportional to a positive power of kinematic invariants in Planck units, since Newton's constant $G_\text{N}$ has negative mass dimension.}, \emph{a priori} one has
\begin{eqaed}
    \Lambda_\text{UV} \lesssim \Lambda_\text{QG} \lesssim \Lambda_\sp \lesssim M_\text{Pl}^{(d)} \, .
\end{eqaed}
According to the original considerations in \cite{Dvali_2010, Dvali:2007wp, Dvali:2010vm, Dvali:2014ila}, and as more convincingly established in \cite{Bedroya:2024ubj}, the UV cutoff also coincides with the parametric curvature scale of the smallest (possibly higher-dimensional) black hole in the theory. An additional EFT-based argument to see this is that quantities such as the entropy afford an asymptotic expansion in inverse powers of the cutoff and horizon area $A$, whose resummation in the full theory should have the schematic form
\begin{eqaed}
    S = \frac{A}{4\ell_\text{Pl}^2} \, f(\Lambda_\text{UV}^2 A) 
\end{eqaed}
for some dimensionless function $f$. Since there are no additional parameters, for areas of the order of the UV cutoff $f(\mathcal{O}(1)) = \mathcal{O}(1)$, while for parametrically smaller areas there can be significant parametric deviations from the Bekenstein-Hawking result. Similar considerations hold for \emph{e.g.} the mass. The key observation is that the leading-order result is driven by the Planck scale, rather than the UV cutoff (whenever they differ). Because of this identification, it has been suggested \cite{Cribiori:2023ffn} that a more instructive quantity to consider is the state counting implemented by the entropy of minimal black holes $S_\text{sp}$, rather than the mere number of species, since the latter may fail to be well-defined outside of infinite-distance limits. \\

Up to the introduction of extra dimensions, decompactification limits seem to have $\Lambda_\text{UV} = \Lambda_\text{QG} = \Lambda_\sp = M_\text{Pl}^{(d+n)}$, at least for toroidal internal spaces where no curvature spoils the analysis. Without introducing extra dimensions explicitly, augmenting the EFT with a self-consistent subset of Kaluza-Klein modes raises the cutoff to $\Lambda_\sp$ by the parametric self-consistency of \cref{eq:speciesdef}, where any EFT irredeemably breaks down and no effective description of black holes is available \cite{Bedroya:2024ubj}, and this is the setting we focus on in the remainder of this section. In this context, \emph{a priori} the cutoff may have still been lower than $\Lambda_\sp$, but it does not appear to be the case. As emphasized in \cite{Basile:2023blg}, this is a non-trivial observation, since for weakly coupled critical strings the UV cutoff $\Lambda_\text{UV} = M_s$ of the EFT is the string scale, while the ``counting'' scale $\Lambda_\sp = M_s \log g_s^{-2}$ \cite{Castellano:2022bvr} bounds the (estimated) strong quantum gravity scale $\Lambda_\text{QG} = M_s \sqrt{\log g_s^{-2}}$ found by studying high-energy string scattering \cite{Gross:1987kza, Gross:1987ar, Dvali:2014ila, Bedroya:2022twb} and its fixed-angle resummation estimates \cite{Mende:1989wt}. Hence, it is worth asking whether internal curvatures can spoil the equality and the scaling $\Lambda_\sp = m_\text{gap}^{\frac{n}{n+d-2}}$ for $n$ decompactifying extra dimensions with a Kaluza-Klein gap $m_\text{gap} \ll \Lambda_\sp$. \\

In the following we consider an initial $d$-dimensional theory as a dimensional reduction of a $(d+n)$-dimensional theory over a $n$-dimensional internal (closed, Riemannian) manifold $X$. In the large-volume limit, as anticipated, the spectrum has a parametrically high number of Kaluza-Klein modes below any fixed energy scale, signaling the breakdown of the ``strict'' $d$-dimensional effective theory cutoff at $m_\text{gap}$. In this setting, in order to study both $\Lambda_\sp$ and $\Lambda_\text{UV}$ it is very convenient to employ the (diagonal, local) heat kernel expansion \cite{Vassilevich_2003}, which we briefly review in \cref{app:heat}, applied to the Laplacian operator $\Delta$ of the internal manifold. The quantity of interest is the heat kernel (trace), defined as
\begin{eqaed}
    K_{X}(t,f) \equiv \mathrm{Tr}_{X}\left[f \, e^{-t\Delta}\right],
\end{eqaed}
where $f \in C^{\infty}(X)$. Since in our case we are interested in considering the heat kernel on the whole manifold, we take $f = 1$ in what follows and denote the resulting trace by $K_X(t)$. Introducing a basis of eigenfunctions of the Laplace operator (\emph{i.e.} Kaluza-Klein modes), this simply reads
\begin{eqaed}
    \mathrm{Tr}_{X}\left[e^{-t\Delta}\right] = \sum_{k}e^{-t\lambda_k},
\end{eqaed}
where $\lambda_k \propto m^2_k$ denotes the eigenvalue of the Laplacian and coincides with the squared mass of the (possibly degenerate) $k$-th excitation level. Notice that the value $K_{X}(0)$, which simply counts the number of modes, is divergent. One could view the definition in \cref{eq:speciesdef} as the sharp cutoff regularization of this divergent sum, namely
\begin{eqaed}
    N_\sp  = \sum_m H\left(1-\frac{m}{\Lambda_\sp}\right),
\end{eqaed}
where $H(x)$ is the Heaviside step function. In order to connect these ideas more directly with the heat kernel and its asymptotic expansion(s), it is first worth exploring the possibility of expressing the number of species as the infinite sum appropriately regularized with a generic smooth function $\eta:\mathbb{R}^+ \rightarrow \mathbb{R}$ rapidly decaying outside of $[0,1]$:
\begin{eqaed}
     N_\sp^{\text{smooth}}  \equiv \sum_m \eta\left(\frac{m}{\Lambda_\sp}\right) .
\end{eqaed}
In other words, we are interested in comparing the asymptotic behavior of the partial sums and of the smooth sums, in the limit where $N_\sp \gg 1$. To do so, we follow an argument by Tao \cite{Tao11}, extended to Schwartzian functions in \cite{Padilla:2024mkm}, which we present in \cref{app:tao}. The upshot is that the two definitions yield the same leading-order answer up to a multiplicative constant (thus parametrically the same), \emph{regardless} of the choice of $\eta$ in the appropriate class. \\

In particular, this means that one can employ an exponential cutoff in the sum \eqref{eq:speciesdef}, which certainly obeys \eqref{eq:rapidvanish}, yielding an expression of the number of species in terms of the heat kernel:
\begin{eqaed}\label{eq:kernelspecies}
    K_X(t=\Lambda_\sp^{-2}) = \sum_m \exp\left(-\frac{m^2}{\Lambda_\sp^2}\right) \equiv N_\sp^{(\text{exp})} = \Lambda_\sp^{2-d}.
\end{eqaed}
The advantage of this definition is that it is independent on the behavior of $K_X$ strictly at $t=0$. Instead, in Planck units, in the limit $m_\text{gap} \propto V^{-\frac{1}{n}} \ll 1$ where the internal volume $V$ grows large, one has $m_\text{gap} \ll \Lambda_\text{sp} \ll 1$ and we can employ the know asymptotic expansion of the heat kernel trace in the $t \rightarrow 0$ limit to extract information on the species scale. Indeed, in this limit the Seeley-DeWitt local diagonal expansion tells us that
\begin{eqaed}
    K_{X}(t) \sim \frac{1}{(4\pi t)^{\frac{n}{2}}}\sum_{k=0}^{\infty} a_{2k}^X \, t^k,
\end{eqaed}
where $a_{2k}^{X}$ are the (even) Seeley-DeWitt coefficients defined and reviewed briefly in \cref{app:heat}. The leading one encoded Weyl's volume law,
\begin{eqaed}
    K_{X}(t) = \frac{V}{(4\pi t)^{\frac{n}{2}}} + \mathcal{O}(t^{\frac{2-n}{2}}) \, ,
\end{eqaed}
and thus for $n$ decompactifying internal dimensions 
\begin{eqaed}
    \Lambda_\text{sp}^{2-d} \sim \frac{\Lambda_\text{sp}^n}{m_\text{gap}^n} \quad \longrightarrow \quad \Lambda_\text{sp} \sim m_\text{gap}^{\frac{n}{n+d-2}} \equiv \Lambda_\text{sp}^{(0)} \, ,
\end{eqaed}
recovering the standard result for tori as expected. However, the identification in \cref{eq:kernelspecies} allows one to go beyond the leading term and compute corrections to the (counting) species scale using the (local, diagonal) heat kernel expansion, which we briefly review in \cref{app:heat}. For instance, if the internal manifold has non-vanishing (mean) scalar curvature $\overline{R} \equiv \frac{1}{V}\int \star R$, one obtains
\begin{eqaed}
    \Lambda_\text{sp}^{2-d} \sim \frac{\Lambda_\text{sp}^n}{m_\text{gap}^n} \left(1 + \frac{\overline{R}}{6\Lambda_\text{sp}^2} \right) \quad \longrightarrow \quad \Lambda_\text{sp} \sim \Lambda_\text{sp}^{(0)} \left(1 - \frac{1}{6(n+d-2)} \, \frac{\overline{R}}{\Lambda_\text{sp}^{(0)}} \right) .
\end{eqaed}
In general the first non-vanishing correction will involve some (local) curvature invariant. Since $\Lambda_\text{sp}^{(0)}$ is the $(d+n)$-dimensional Planck scale, and the higher-dimensional effective theory is reliable only for typical curvatures that are parametrically subplanckian, the above correction can never become significant within the regime of validity of the setup. Therefore, in order to seek significant deviations from this geometric regime, one has to look elsewhere. \\

Two natural avenues to this end are the study of moduli spaces from the worldsheet point of view in geometric settings, where stringy effects may play a role, and (possibly) non-geometric settings, where the pattern dictated by the emergent string conjecture may fail to hold. In the following section, as well as in \cref{app:geometric_vacua}, we show that the former case provides no qualitative differences, confirming that a worldsheet approaches is compatible with the EFT approach in geometric settings. In the latter case, presented in detail in \cref{sec:species_worldsheet}, we shall show that even without assuming geometric points in the conformal manifold of the worldsheet CFT, a proxy for the UV cutoff and the species scale vanishes if and only if a light tower of species appears. Furthermore, under suitable assumptions stemming from the swampland distance conjecture \cite{Ooguri:2006in}, it does indeed behave in a geometric-like fashion with the spectral gap, indicating the possible emergence of geometry which is a crucial ingredient for the emergent string conjecture. Before proceeding in this direction, we examine the UV cutoff $\Lambda_\text{UV}$ with the heat kernel approach, refining the argument in \cite{Basile:2023blg} by which decompactification limits have $\Lambda_\text{UV} = \Lambda_\sp$ by addressing a subtlety in the EFT calculation.

\subsection{Endangered species scale and the ultraviolet cutoff}\label{sec:endangered}

From the above analysis we learned that, in controlled EFT compactifications with $m_\text{gap} \ll 1$, the species scale $\Lambda_\sp \sim \Lambda_\sp^{(0)} \equiv m_\text{gap}^{\frac{n}{n+d-2}} = M_\text{Pl}^{(d+n)}$. In addition to counting Kaluza-Klein modes, the heat kernel is also particularly convenient to compute one-loop effective actions, which contain information about the UV cutoff $\Lambda_\text{UV}$ of the EFT. As discussed in \cite{Basile:2023blg}, the relevant sector of the effective action contains tree-level terms which are suppressed by the higher-dimensional Planck scale, $\Lambda_\text{UV}^\text{tree-level} = \Lambda_\sp$, while one-loop contributions to the quantum effective action can be estimated from dimensional analysis. The (local, diagonal) heat kernel expansion, which we briefly review in \cref{app:heat}, is a powerful tool to present the result in a manifestly covariant fashion. The upshot is that the one-loop terms of the effective action can be recast in an asymptotic series of the form
\begin{eqaed}
    S_\text{1-loop} \sim -\,\frac{1}{2(4\pi)^{\frac{d}{2}}}\int_{\Lambda_\sp^{-2}}^\infty \frac{dt}{t^{1+\frac{d}{2}}} \, K_X(t) \sum_{k \geq 0} a_{2k}(\mathcal{R}) \, t^k \, ,
\end{eqaed}
where $\mathcal{R}$ collectively denotes Riemann curvatures and covariant derivatives thereof. In the decompactification limit, as we have seen $K_X$ is dominated by $m_\text{gap}^{-n} t^{-\frac{n}{2}}$ and thus, performing a change of variables, the coefficients of dimension $2k$ operators with $2k < d+n$ are dominated by
\begin{eqaed}
    m_\text{gap}^{d-2k} \int_{\frac{m_\text{gap}^2}{\Lambda_\sp^2}} \frac{ds}{s} \, s^{k-\frac{d+n}{2}} \sim m_\text{gap}^{d-2k} \left( \frac{\Lambda_\sp}{m_\text{gap}}\right)^{d+n-2k} = \Lambda_\sp^{2-2k} \, ,
\end{eqaed}
while the infrared integration limit is tamed by the exponential decay of the heat kernel at large $t$. This further supports the tree-level identification $\Lambda_\text{UV} = \Lambda_\sp$. On the other hand, operators with $2k \geq d+n$ are not accompanied by divergent scalings of this type, and thus naively it would seem that their coefficients are controlled entirely by $m_\text{gap}$ rather than $\Lambda_\sp$, without any appearance of the Planck scale. In other words, these operators appear to contribute with a $m_\text{gap}$ suppression ``additively'' relative to the Planck-sensitive couplings, according to $\delta S_\text{eff} \propto \int m_\text{gap}^{d-2k} \, a_{2k}(\mathcal{R})$, rather than ``multiplicatively'' as in terms of the form $\delta S_\text{eff} \propto M_\text{Pl}^{d-2} \int \Lambda_\text{UV}^{2-2k} \, a_{2k}(\mathcal{R})$. These schemata suggest an interpretation along the lines of \cite{Bedroya:2024ubj} by which the low-energy expansion of observables roughly takes the form
\begin{eqaed}
    \mathcal{A}_\text{QG}(E) \sim \mathcal{A}_\text{GR}(E) \, f\left(\frac{E}{\Lambda_\text{UV}}\right) + \mathcal{A}_\text{KK}\left(\frac{E}{m_\text{gap}}\right) ,
\end{eqaed}
where the latter term can be accounted for augmenting the theory with Kaluza-Klein species, while the former is a genuine quantum gravity effect which also survives decompactifications. While the latter type of suppression is likely to occur for some operators, at least in the ``strict'' EFT with the whole Kaluza-Klein spectrum integrated out (for instance those implementing Schwinger production), the above reasoning may not suffice, since such one-loop terms would dominate over tree-level terms, indicating that within the EFT this expansion is not under control. The appearance of other scales in between $m_\text{gap}$ and $\Lambda_\sp$ would indicate a different cutoff for the EFT augmented with Kaluza-Klein spectra, where the effects suppressed by $m_\text{gap}$ are ``integrated in'', as \emph{e.g.} in \cite{Bedroya:2024ubj}.

\subsection{String perturbation theory to the rescue}\label{sec:stringy_rescue}

To see how this puzzle is resolved in string theory, let us consider the one-loop $R^4$ terms in type II toroidal compactifications\footnote{This setting allows for exact computations via (U-)dualities and BPS protection \cite{Green:1997as, Green:1997di, Kiritsis:1997em, Pioline:1997pu, Obers:1998fb, Green:2010sp, Pioline:2010kb}, a remarkable property which was recently revisited in \cite{Blumenhagen:2023xmk, Blumenhagen:2024lmo, Blumenhagen:2024ydy}. With the generalization to \cref{sec:species_worldsheet} in mind, we restrict ourselves to one-loop level in string perturbation theory.} \cite{Green:1982sw, Obers:1999um, Green:1999pv, Green:2008uj, Green:2010wi, Angelantonj:2011br, Angelantonj:2012gw, Angelantonj:2013eja}. Their Wilson coefficients are proportional to (regularized) modular integrals of Narain lattice sums $\Gamma_{n,n}$, namely integrals over a fundamental domain $\mathcal{F}$ of the moduli space of the worldsheet torus,
\begin{eqaed}
    c^{\text{1-loop}}_\text{Narain} = \int_{\mathcal{F}}d\mu \, \Gamma^{\text{reg}}_{n,n}
\end{eqaed}
with $d\mu \equiv \frac{d^2\tau}{(\Im \tau)^2}$ the modular invariant measure of volume $\frac{\pi}{3}$, and a regularization which we shall introduce in detail in \cref{sec:species_worldsheet}. The exact expression for the renormalised integral of the lattice term has been computed in several ways \cite{Obers:1999um, Green:2010wi, Angelantonj:2012gw}, and results in an $SO(n,n)$ Langland-Eisenstein series with weight vector $[1,0^{n-1}]$\footnote{The case $n=1$ can be treated separately, yielding the exact result $I_1 = \frac{\pi}{3}(r+r^{-1})$ for a circle of radius $r$ in string units. For the purposes of this paper, since unitary $c=1$ CFTs (with a vacuum state \cite{Runkel:2002yb}) are geometric \cite{Dijkgraaf:1987vp}, this case can be neglected.}.
\begin{eqaed}
    I_n= \frac{\Gamma\left(\frac{n}{2}-1\right)}{\pi^{\frac{n}{2}-1}} \, E^{SO(n,n)}_{[1,0^{n-1}],\frac{n}{2}-1} \, . 
\end{eqaed}
The large-volume decompactification limit $\mathcal{V} \equiv \alpha'^{-\frac{n}{2}} V_n \gg 1$ leads to \cite{Obers:1999um, Green:2010wi, Angelantonj:2011br}
\begin{eqaed}\label{eq:LanglEisSO}
    E^{SO(n,n)}_{[1,0^{n-1}],\frac{n}{2}-1} \sim \mathcal{V} \, \frac{\pi^{\frac{n}{2}}}{3 \Gamma\left(\frac{n}{2}-1\right)} + \mathcal{V}^{1-\frac{2}{n}} \, E^{SL(n)}_{[0^{n-2},1],\frac{n}{2}-1} \, .
\end{eqaed}
The latter series satisfies the recursion relation \cite{Green:2010wi}
\begin{eqaed}
    E^{SL(n)}_{[0^{n-2},1],\frac{n}{2}-s}= \frac{\Gamma(s)}{\pi^{2s-\frac{n}{2}} \Gamma\left(\frac{n}{2}-s\right)} \, E^{SL(n)}_{[1,0^{n-2}],s} \, ,
\end{eqaed}
from which one can extract the analytic properties from those of $E^{SL(n)}_{[1,0^{n-2}],s}$. It is known that for $n \geq 3$ the latter is regular at $s=1$, implying that $E^{SL(n)}_{[0^{n-2},1],\frac{n}{2}-1}$ is regular as well. An issue may arise for $n=2$ where $E^{SL(n)}_{[1,0^{n-2}],s}$ has a pole at $s=1$. However, the Laurent expansion is known and compensates the pole of the Euler gamma function. The upshot is that $E^{SL(n)}_{[0^{n-2},1],\frac{n}{2}-1}$ is regular also for $n=2$, and thus \cref{eq:LanglEisSO} corresponds to the asymptotic behavior found previously. \\

Instead of an isotropic large-volume limit to the critical dimension $d+n=10$, let us consider decompactifications from $d=7$ to $d+n=8$, where the EFT analysis becomes marginally reliable for $R^4$ terms. In the limit in which the radius $r$ (in string units) of a singled-out circle inside $T^3 = T^2_\text{fixed} \times S^1_{r}$ is large, the Wilson coefficients scales as a linear combination of $r$ and $r \, \log(r)$ \cite{Green:2010wi, Angelantonj:2011br, Benjamin:2021ygh}. The latter logarithm reflects the marginal reliability of the EFT calculation, and it appears in eight dimensions because of the dimensionless threshold terms associated to the $R^4$ operator, and we shall discuss its appearance in general worldsheet CFTs in \cref{sec:species_worldsheet}. The power-like scaling is given by the volume of the decompactifying space, which leads to a species-scale suppression. To see this, one can observe that the corresponding term in the (Einstein-frame\footnote{For decompactification limits of this type the dilaton dependence is not relevant, and $M_s = M_\text{Pl}^{(d+n)} = M_\text{Pl}^{(10)}$ parametrically. We will include it in the more general analysis of \cref{sec:species_worldsheet}.}) effective action is
\begin{eqaed}
    (M_\text{Pl}^{(d+n)})^{d-2}\int d^dx \sqrt{-g} \, (M_\text{Pl}^{(d+n)})^n \, m_\text{gap}^{-n} \, \frac{t_8t_8R^4}{(M_\text{Pl}^{(d+n)})^6} \, ,
\end{eqaed}
where $t_8t_8R^4$ denotes a particular dimension-eight operator quartic in the Riemann curvature \cite{Green:1997di}. Thus, using $ (M_\text{Pl}^{(d+n)})^{d+n-2} = (M_\text{Pl}^{(d)})^{d-2} m_\text{gap}^n$, the coefficient in $d$-dimensional Planck units is given by
\begin{eqaed}
    m_\text{gap}^{-n} \times m_\text{gap}^{\frac{n(d+n-8)}{d+n-2}} = \Lambda_\sp^{-6} 
\end{eqaed}
as expected. Thus, whenever the Wilson coefficient $c^{\text{1-loop}}$ scales like a volume in string units, the resulting suppressing scale is the species scale, except for the threshold logarithm. This result is consistent with the notion that, indeed, for \emph{bona fide} decompactification limits the UV cutoff $\Lambda_\text{UV}$ of the gravitational theory (augmented by Kaluza-Klein species up to $\Lambda_\sp$) agrees with the ``counting'' species scale $\Lambda_\sp$ as originally defined in \cite{Veneziano:2001ah, Dvali_2010}. \\

To see what happens when $R^4$ terms are strictly outside the validity of the heat-kernel EFT computation, we now consider $S^1_r$ decompactifications to dimensions $d+n < 8$, with $n=1$. As shown in \cite{Angelantonj:2011br}, and as we shall revisit in more detail in \cref{sec:factorized_limits} in more general settings, in addition to the expected term linear in the radius $r$ of the large circle, there is an additional term of the type $r^{k-1}$, with $k=9-d$ the number of leftover compact dimensions. Hence, if $d+1<8$ this term dominates. As we shall see in more generality in \cref{sec:factorized_limits}, the corresponding term in the effective action schematically reads $\delta S_\text{eff} = \int m_\text{gap}^{d-8} \, R^4$. In other words, these ``superleading'' terms are precisely of the ``additive'' type as discussed in the preceding subsection. A similar candidate in ten dimensions is the $D^4R^4$ operator studied in \cite{Green:1999pu}\footnote{We are grateful to A. Castellano and A. Herr\'{a}ez for discussions on this point.}. \\

We shall now extend the above discussion to potentially non-geometric settings, showing how the swampland distance conjecture can be derived assuming that the UV cutoff be parametrically small in Planck units, and then conversely how the same volume-like scalings appear under some assumptions on the internal CFT with a small gap. Concretely, in \cref{sec:light_tower} we derive a specific form of the swampland distance conjecture from the assumption that the UV cutoff be small in Planck units, whereas in \cref{sec:non-factorized_limits} and \cref{sec:factorized_limits} we study such infinite-distance limits, recovering the same geometric behavior for one-loop $R^4$ corrections that we have discussed in \cref{sec:stringy_rescue}, namely that the UV cutoff is the higher-dimensional Planck scale $\Lambda_\text{sp}$. In order to conceptually connect this approach with the one presented in the preceding section, in \cref{app:geometric_vacua} we introduce a systematic method to define and compute distances in geometric worldsheet vacua in the presence of curvature, matching them to the analogous EFT quantities along the lines of \cite{Lust:2019zwm}.

\section{Emergent geometry from worldsheet CFTs}\label{sec:species_worldsheet}

In this section we extend the computations of one-loop $R^4$ terms in type II vacua \cite{Green:1982sw, Obers:1999um, Green:1999pv, Green:2008uj, Green:2010wi, Angelantonj:2011br, Angelantonj:2012gw, Angelantonj:2013eja} to potentially non-geometric settings, described by a generic internal CFT which we take to be unitary, critical, compact, modular invariant and with a unique vacuum. Analogous considerations can be made for $R^2$ terms \cite{Gregori:1997hi}. To begin with, we discuss the computation of the relevant Wilson coefficient, and then we turn to its moduli dependence in the following (sub)sections.

\subsection{Higher-derivative corrections from graviton scattering}\label{sec:graviton_scattering}

In type II backgrounds\footnote{More generally, our formalism applies to RNS-RNS constructions for purely closed strings. Since type 0 projections of this type generally contain physical tachyons, we shall restrict to type II to avoid classical instabilities. In general, classical stability in these settings requires the presence of spacetime fermions \cite{Angelantonj:2023egh}, while for open strings the story is more complicated \cite{Cribiori:2020sct, Leone:2023qfd}.}, the four-graviton amplitude $\mathcal{A} \equiv \mathbf{K} \, A(s,t,u)$ can be written in terms of a universal kinematic factor $\mathbf{K}$ containing external momenta and polarization contractions, which can be derived in Einstein gravity, and a Bose-symmetric function $A$ of the kinematic Mandelstam variables $s \, , \, t \, , \, u$. The amplitude receives no contributions from the internal sector at tree level. At one loop, the torus correlator of graviton vertex operators is accompanied by the partition function of the internal CFT. Multiplying and dividing by its flat-space counterpart to restore the critical flat-spacetime integrand, the total amplitude is modified according to \cite{Green:1982sw}
\begin{eqaed}
    A_\text{1-loop} = \int_\mathcal{F} d\mu \, F(s,t,u|\tau, \overline{\tau}) \quad \longrightarrow \quad \int_\mathcal{F} d\mu \, F(s,t,u|\tau, \overline{\tau}) \, \frac{Z_\text{int}(\tau, \overline{\tau})}{Z_\text{ext}(\tau, \overline{\tau})} \, ,
\end{eqaed}
where $F$ is a critical spacetime factor containing the kinematic data and worldsheet propagators on the torus, whereas $Z_\text{int}$ is the torus partition function of the internal worldsheet CFT and $Z_\text{ext}$ is its counterpart when the internal sector is replaced by flat space. Namely, its bosonic sector is $(\sqrt{\Im \tau}\abs{\eta(\tau)}^2)^{-c}$. For Narain theories, the result of this ratio is the primary partition function of the $U(1)^c \times U(1)^c$ current algebra, namely the lattice sum $\Gamma_{n,n}$. More generally, the contribution of fermionic (super)conformal characters and sum over spin structures amounts to a suitable helicity supertrace \cite{Gregori:1997hi}, which we absorb in this ratio defining the ``reduced'' partition function \cite{Afkhami-Jeddi:2020hde}
\begin{eqaed}\label{eq:reduced_partition_function_def}
    \mathcal{Z} \equiv \frac{Z_\text{int}}{Z_\text{ext}} = y^\frac{c}{2} \abs{\eta}^{2c} Z'_\text{int} = y^{\frac{c}{2}} \sum_{j , \Delta} e^{2\pi i j x} e^{-2\pi \Delta y} \, ,
\end{eqaed}
where, here and in the following, we suppress the $\tau \, , \overline{\tau}$ argument and we denote explicit factors of $\tau$ as $\tau = x + iy$. One can think of such internal partition function stripped of fermionic characters in terms of a CFT of central charges $c_L = c_R \equiv c = 10-d$. \\

The low-energy expansion of the amplitude derives from the expansion of $F$, which contains analytic and non-analytic terms \cite{Green:1999pv, Green:2008uj}. Although the amplitude is finite, splitting the integrand to isolate analytic terms can produce spurious divergences. Since we are ultimately interested in the moduli dependence of the analytic terms, a suitably regularization procedure \`{a} la Rankin-Selberg \cite{Rankin1939ContributionsTT, selberg1940bemerkungen, BUMP198949, Don1982TheRM, Angelantonj:2010ic, Angelantonj:2011br, Benjamin:2021ygh} can be used to correct for expanding $F$ under the integral sign. Since gravitons satisfy $s+t+u=0$, no $R^2$ corrections can appear in the amplitude because $\frac{1}{st} + \frac{1}{tu} + \frac{1}{us} = 0$, while the $R^3$ correction proportional to $\frac{1}{s}+\frac{1}{t}+\frac{1}{u}$ does not appear since $F(0,0,0|\tau,\overline{\tau}) = 1$. Thus, the $R^4$ correction is the first to arise from this amplitude, and its one-loop Wilson coefficient is thus given by
\begin{eqaed}
    c^\text{1-loop}_{R^4} = \int_\mathcal{F} d\mu \, \mathcal{Z}^\text{reg} \, .
\end{eqaed}
As an example, in ten-dimensional type IIA string theory the total (and one-loop exact) coefficient in the amplitude normalized as in \cite{Guerrieri:2021ivu, Guerrieri:2022sod} relative to $\frac{1}{stu}$ is
\begin{eqaed}
    \alpha_\text{IIA}^{10d} = g_s^{\frac{1}{2}} \, \frac{\pi^2}{96} + \frac{\zeta(3)}{32g_s^{\frac{3}{2}}} = \frac{\sqrt{g_s}}{64} \left( \frac{1}{g_s^2} \, 2\zeta(3) + g_s^0 \,  2\pi \, \frac{\pi}{3}\right).
\end{eqaed}
This decomposition matches the string-frame conventions of \cite{Green:1999pv}, whereby the tree-level amplitude is 
\begin{eqaed}
    A_\text{string}(s,t,u) = 64\pi^7 \ell_\text{Pl}^8 \left( \frac{64}{{\alpha'}^3stu} + 2\zeta(3) + \dots \right) .
\end{eqaed}
The total amplitude in \cite{Guerrieri:2021ivu} is instead normalized according to Planck units,
\begin{eqaed}
    A_\text{Planck}(s,t,u) = 64\pi^7 \ell_\text{Pl}^8 \left(\frac{1}{stu} + \alpha \, \ell_\text{Pl}^6 + \dots \right) ,
\end{eqaed}
and since $\ell_\text{Pl}^8 = g_s^2 \ell_s^8$ this identifies the structure
\begin{eqaed}
    A(s,t,u) = \frac{(2\pi)^7\ell_s^8}{2} \, g_s^4 \left( \frac{1}{g_s^2} \, \frac{1}{stu} + g_s^{-\frac{1}{2}} \, \ell_s^6 \, \alpha + \dots \right)
\end{eqaed}
carried by the properly normalized vertex operators and string perturbation theory. The above expression shows that $\alpha/\sqrt{g_s}$ is the string-frame Wilson coefficient. Indeed, since $g_s$ is the zero-mode of the dilaton and does not enter the change of frame, the string-frame genus-$g$ effective action term
\begin{eqaed}
    M_s^8\int d^{10}x \, \sqrt{-g_S} \, e^{(2g-2)\phi} \alpha_g^{(\text{s})} \, \frac{\mathcal{O}_4(R)}{M_s^6} = M_\text{Pl}^8 \int d^{10}x \, \sqrt{-g} \, \alpha_g^{(\text{s})} \, g_s^{2g-2+\frac{1}{2}} \, \frac{\mathcal{O}_4(R)}{M_\text{Pl}^6}
\end{eqaed}
for the quartic curvature operators yields an Einstein-frame Wilson coefficients $\alpha_g \propto \sqrt{g_s} \, g_s^{2g-2}\alpha_g^{(\text{s})}$. All in all, in general due to the extra $2\pi$ in the one-loop amplitude, for type II theories one has
\begin{eqaed}\label{eq:proxy_cutoff}
    \alpha_\text{II}^\text{1-loop} = \frac{\sqrt{g_s}}{64} \left(\frac{2\zeta(3)}{g_s^2} + 2\pi \, c_{R^4}^{\text{1-loop}} \right) .
\end{eqaed}
In this context, since we are dealing with weakly coupled strings, unsurprisingly one can straightforwardly recover part of the dichotomy of infinite-distance limits. To wit, when the former term dominates ($g_s \ll 1$ at fixed moduli) the associated cutoff
\begin{eqaed}
    \frac{\Lambda_\text{UV}}{M_\text{Pl}} = \abs{\alpha}^{-\frac{1}{6}} \overset{g_s \to 0^+}{\sim} g_s^\frac{1}{4} = \frac{M_s}{M_\text{Pl}}
\end{eqaed}
yields the string scale $M_s$ as expected. When the latter term dominates, one must also identify the proper Planck scale $M_\text{Pl}$ of the EFT relative to the Planck scale of the critical flat-spacetime EFT $M_{10}$. As we have discussed in \cref{sec:species_compactifications}, if $c_{R^4}^{\text{1-loop}}$ scales ``as a volume'' (in a sense which we shall make precise in the following), the (proxy for the) UV cutoff $\Lambda_\text{UV}$ coincides with the species scale $\Lambda_\sp$ for decompactifications. Generally, the precise cutoff scale should be evaluated taking all higher-derivative corrections into account, in the spirit of \emph{e.g.} \cite{vandeHeisteeg:2023dlw}. However, it is natural to expect Wilson coefficients to be of the same parametric order, as supported by analyses of EFT positivity bounds (see \emph{e.g.} \cite{Arkani-Hamed:2020blm, deRham:2022hpx}), up to (likely finitely many \cite{Heckman:2019bzm}) fine tunings. Therefore, we expect that the quartic curvature correction provides a reliable proxy for the UV cutoff of the gravitational sector, and that \cref{eq:proxy_cutoff} be valid at least when the one-loop calculation is reliable. In principle one can choose almost any gravitational higher-derivative correction, some exceptions including topological terms induced by the Green-Schwarz mechanism \cite{vandeHeisteeg:2023dlw}. The quartic term we focus on has the advantages of surviving maximal supersymmetry and being well-studied in many examples in the literature \cite{Green:1982sw, Obers:1999um, Green:1999pv, Green:2008uj, Green:2010wi, Angelantonj:2011br, Angelantonj:2012gw, Angelantonj:2013eja} also beyond the one-loop approximation \cite{Green:1997as, Green:1997di, Kiritsis:1997em, Pioline:1997pu, Obers:1998fb, Green:2010sp, Pioline:2010kb, Blumenhagen:2023xmk, Blumenhagen:2024lmo, Blumenhagen:2024ydy}. \\

Let us first recall how the story goes in geometric compactifications, as discussed in \cref{sec:species_compactifications}. The tree-level term is insensitive to the compactification and thus it contributes like a ten-dimensional term evaluated on the compactified background, yielding
\begin{eqaed}
    M_\text{Pl}^{d-2}\int d^dx \, g_s^{-2} \, 2\zeta(3) \, t_8t_8 R^4
\end{eqaed}
in string units and string frame. The one-loop term is evaluated directly in the $d$-dimensional theory, leading to
\begin{eqaed}
    \int d^dx \, 2\pi \, c_{R^4}^{\text{1-loop}} \, t_8t_8 R^4
\end{eqaed}
in string units and string frame \cite{Green:1997di}. Keeping track of the change of variables to the Einstein frame and collecting a volume factor
\begin{eqaed}
    \frac{V_{10-d}}{\ell_s^{10-d}} \equiv \frac{M_s^{10-d}M_\text{Pl}^{d-2}}{M_{10}^8} = g_s^2 \left(\frac{M_\text{Pl}}{M_s}\right)^{d-2} ,
\end{eqaed}
all in all in type II one ends up with the Einstein-frame action
\begin{eqaed}
    S_\text{eff} \propto \frac{M_\text{Pl}^{d-2}}{2} \int d^dx \, \sqrt{-g} \, \alpha \, \frac{t_8t_8 R^4}{M_\text{Pl}^6} \, ,
\end{eqaed}
where the Wilson coefficient $\alpha \propto \left(\frac{M_\text{Pl}}{\Lambda_\text{UV}}\right)^6$ appearing in the above amplitude as $\frac{1}{stu} + \alpha \, \ell_\text{Pl}^6$ reads
\begin{eqaed}
    \alpha & = \frac{\sqrt{g_s}}{64} \left(\frac{M_\text{Pl}}{M_{10}}\right)^6 \left( \frac{2\zeta(3)}{g_s^2} + \frac{2\pi}{g_s^2} \left(\frac{M_s}{M_\text{Pl}}\right)^{d-2} \int_\mathcal{F} d\mu \, \mathcal{Z}^{\text{reg}} \right) \\
    & = 2^{-6} \left(\frac{\ell_s}{\ell_\text{Pl}}\right)^{8-d} \left(2\zeta(3) \left(\frac{\ell_s}{\ell_\text{Pl}}\right)^{d-2} + 2\pi \int_\mathcal{F} d\mu \, \mathcal{Z}^{\text{reg}} \right)
\end{eqaed}
with the normalization of \cite{Guerrieri:2021ivu, Guerrieri:2022sod}. This expression reproduces the correct result in toroidal compactifications, where large-volume limits yield the expected scaling $\Lambda_\text{UV} = \Lambda_\sp$ for decompactification limits. Specifically, when the first term dominates, $\alpha \sim 2^{-6} \, 2\zeta(3) \, \left(\frac{\ell_s}{\ell_\text{Pl}}\right)^6$ yields $\frac{\Lambda_{\text{UV}}}{M_\text{Pl}} = \frac{M_s}{M_\text{Pl}}$ independently of the dimension, as we have already seen. When the latter term dominates, the large-volume limit discussed in \cref{sec:species_compactifications} yields $\Lambda_\text{UV} = m_\text{gap}^{\frac{10-d}{10-2}}$ as expected from a decompactification from $d$ to 10 dimensions. Similarly, for partial decompactifications one finds the expected result. This suggests that the interplay of limits we consider yields a physically relevant scale from the one-loop computation; a resummed estimate along the lines of \cite{Mende:1989wt} could further ground the conclusions we shall draw from this quantity. \\

Let us now generalize these considerations to any internal CFT. The Planck scale $M_\text{Pl}$ should arise from the analogous ratio $\mathcal{Z}_{S^2}$ of sphere partition functions which defines the torus counterpart $\mathcal{Z}_{T^2}$ in \cref{eq:reduced_partition_function_def}. This identification allows to interpret the calculation in purely $d$-dimensional terms. The Weyl anomaly indeed cancels in the ratio, and for $\mathcal{N}=(2,2)$ SCFTs the ratio reproduces the exponentiated K\"{a}hler potential $e^{-K}$ of the conformal manifold\footnote{More precisely, depending on whether type A or type B supersymmetric backgrounds are used for localization on $S^2$, one obtains either chiral or twisted chiral K\"{a}hler potentials.} \cite{Gerchkovitz:2014gta, Gomis:2015yaa}, which furnishes an analog of the internal volume in string units. For example, on square torus of radius $R$ one has $K=-n \log R$, and thus $e^{-K} \propto \mathcal{V}$ as expected. \\

In general, one thus has $M_\text{Pl}^{d-2} = M_{10}^8 \, \mathcal{Z}_{S^2}$ in string units, or in other words the one-loop Wilson coefficient for any internal CFT is given by
\begin{eqaed}\label{eq:final_wilson_coeff}
    \alpha_\text{II} = 2^{-6} \left( \frac{\mathcal{Z}_{S^2}}{g_s^2} \right)^{\frac{8-d}{d-2}} \left(2\zeta(3) \, \frac{\mathcal{Z}_{S^2}}{g_s^2} + 2\pi \, \int_{\mathcal{F}} d\mu \, \mathcal{Z}^\text{reg}_{T^2}\right) .
\end{eqaed}
As discussed above, the need to regulate the modular integral for $d<9$ stems from the fact that, while the full one-loop amplitude is finite, the analytic piece of the correction by itself may diverge (as it does for Narain lattices in dimensions $n>1$). The divergence is canceled by a corresponding contribution from the non-analytic part arising from massless loops which is fixed by one-loop unitarization. We shall discuss this aspect in more detail below, where it will play a crucial role for $d=8$ in order to reproduce the threshold corrections to the $R^4$ term. Except for this subtlety, $d=8$ is perhaps the simplest non-trivial case to consider, since the prefactor in \cref{eq:final_wilson_coeff} cancels. As a final comment, let us observe that the ambiguities in the definition of the sphere partition function are carried by Ricci-scalar counterterms \cite{Gomis:2015yaa}, which are precisely those that are reabsorbed by a redefinition of the string coupling $g_s^{\chi} = e^{\chi \langle \phi \rangle} = e^{\langle \phi \rangle \int_{\mathcal{W}} \star R}$, and for the sphere $\chi = -2$. The fact that \cref{eq:final_wilson_coeff} contains only the ratio $\frac{\mathcal{Z}_{S^2}}{g_s^2}$ in all dimensions is thus a non-trivial consistency check. \\

With the above preparations in place, we are now ready to discuss the (regularized) modular integral
\begin{eqaed}
    I(t) \equiv \int_{\mathcal{F}} d\mu \, \mathcal{Z}_{T^2}^{\text{reg}}
\end{eqaed}
in \cref{eq:final_wilson_coeff}, where now $t$ denotes a marginal parameter, expressing $\mathcal{Z}_{T^2}$ as in \cref{eq:reduced_partition_function_def}. For $c>1$ (and thus $c\geq 2$, since we consider integral values $c=10-d$), the reduced partition function is not integrable and is not of rapid decay, but there is a simple way to implement the Zagier improvement \cite{Don1982TheRM, Angelantonj:2010ic, Angelantonj:2011br} of the Rankin-Selberg \cite{Rankin1939ContributionsTT, selberg1940bemerkungen, BUMP198949} method to obtain a ``renormalized'' modular integral without affecting moduli dependence\footnote{Alternative procedures and their connections are discussed in \cite{Angelantonj:2011br}.}. Namely, one can trade the Rankin-Selberg-Zagier transform and the corresponding regulated integral of a suitably slow-growth automorphic function $F(\tau) \overset{y \gg 1}{\sim} y^{\alpha}$ with the ordinary Rankin-Selberg transform and modular integral of the rapidly decaying function $\widetilde{F}(\tau) \equiv F(\tau) - E_\alpha(\tau)$, where $E_\alpha$ denotes a real-analytic Eisenstein series\footnote{For more general growth behavior, one subtracts a suitable linear combination of derivatives of such Eisenstein series \cite{Benjamin:2021ygh}. For our purposes in this paper, the growing contributions spelled out in \cref{eq:reduced_partition_function_def} only include $y^{\frac{c}{2}}$.}. \\

We are thus led to consider integrals of the form
\begin{eqaed}\label{eq:modular_integral_reg}
    I(t) = \int_\mathcal{F} d\mu \, \widetilde{\mathcal{Z}_{T^2}} \, .
\end{eqaed}
More precisely, for $c=2$ one should subtract the regularized (but ``anomalous'') Eisenstein series $\widehat{E}_1$ obtained from the Kronecker limit formula, which indeed also appears in the moduli-dependent $R^4$ corrections in type II string theory on a two-torus. To begin with, we shall show that if $I(t)$ diverges in some limit the gap $\Delta_\text{gap}(t)$ must vanish, bringing along an infinite tower of light states in the reduced partition function. Then, assuming such a scenario, we discuss in detail the scaling with which $I(t)$ diverges.

\subsection{Divergent Wilson coefficient implies light tower}\label{sec:light_tower}

We now show that if $I(t)$ diverges in some limit, the spectrum described by the reduced partition function must feature a light tower of states. To this end, let us assume by way of contradiction that the gap $\Delta_\text{gap}$ does \emph{not} vanish in the limit. Then, adding and subtracting the vacuum contribution $y^{\frac{c}{2}}$, the integrand can be bounded according to
\begin{eqaed}\label{eq:abs_bound}
    \abs{\widetilde{\mathcal{Z}_{T^2}}} \leq y^{\frac{c}{2}}\sum_{j,\Delta>0} e^{-2\pi \Delta y} + \abs{E_{\frac{c}{2}} - y^{\frac{c}{2}}} \, ,
\end{eqaed}
and its modular integral by the integral over the strip $\{ y \geq \frac{\sqrt{3}}{2}\}$. The second term is a constant, which we will neglect in the following. The ``canonical'' reduced partition function $Z \equiv \sum_{j,\Delta} e^{-2\pi \Delta y}$ can be separated into a contribution $Z_{<}$ from states with $\Delta < \Delta_\text{gap}$ and a contribution $Z_{>}$ from states with $\Delta \geq \Delta_\text{gap}$. By construction, $Z_{<} = y^{\frac{c}{2}}$. Following the argument in section 2.2 of \cite{Hartman:2014oaa}, applying inequalities of the form $e^{-2\pi\Delta y} \leq e^{-2\pi \Delta_\text{gap} y}$ to sums over states with $\Delta \geq \Delta_\text{gap}$, modular invariance implies that for $y>1$
\begin{eqaed}\label{eq:light_heavy_bound_1}
    Z_{>} \leq \frac{e^{2\pi \left(\frac{1}{y}-y\right)\Delta_\text{gap}}}{1-e^{2\pi \left(\frac{1}{y}-y\right)\Delta_\text{gap}}} \, Z_{<} \overset{y \gg 1}{\sim} y^{\frac{c}{2}} \, e^{2\pi \left(\frac{1}{y}-y\right)\Delta_\text{gap}} \, ,
\end{eqaed}
whereas, by analogous considerations, for $y < 1$
\begin{eqaed}\label{eq:light_heavy_bound_2}
    Z_{>} \leq \frac{1}{1-e^{2\pi \left(y-\frac{1}{y}\right)\Delta_\text{gap}}} \, Z_{<}^{-1} \, .
\end{eqaed}
Hence, using \cref{eq:light_heavy_bound_1}, the strip integrals of $\sum_{\Delta > 0}e^{-2\pi \Delta y} = Z_{>}$ over the strip $\{y>K>1\}$ is bounded if $\Delta_\text{gap}$ does not vanish. The remaining integral over $\{ \frac{\sqrt{3}}{2} \leq y \leq K\}$ is bounded by $Z\left(y=\frac{\sqrt{3}}{2}\right)$ up to a prefactor. By virtue of \cref{eq:light_heavy_bound_2}, this is in turn bounded by a function of the gap which remains finite. In the general case in which the reduced internal partition function is a linear combination of bosonic characters, analogous bounds and conclusions follow from modular invariance, involving the positive involution matrix which implements the S modular transformation on the characters. \\

The above argument shows that if $I(t)$ diverges with $t$ in some limit the gap must vanish. We can now reiterate the same logic to derive the existence of a light tower of states brought along by the small gap. To see this, let us assume by way of contradiction that finitely many (say $N$) many states become light along with the gap. Then, the remaining states have conformal weights above some new threshold $\Delta_\text{th}$ which remains finite in the limit. Replacing $\Delta_\text{gap}$ by $\Delta_\text{th}$ in the above argument, and bounding $Z_{<} = y^{\frac{c}{2}}\sum_{\Delta \leq \Delta_\text{th}} e^{-2\pi\Delta y} \leq N \, y^{\frac{c}{2}}$, yields a finite bound once more. Hence, there must be an infinite light tower. Combining these results with the recent analysis of \cite{Ooguri:2024ofs} (translated to a worldsheet framework), assuming that the limiting four-point correlators of light operators exist, it follows that the limit lies at infinite distance and that the gap vanishes \emph{exponentially} in the Zamolodchikov distance, with a rate within the interval $[c^{-\frac{1}{2}}, 1]$. This resonates with the considerations in \cite{Stout:2022phm} regarding factorization in the presence of gravity. \\

Let us summarize our findings thus far. Since the universal factorization of infinite-distance limits in gravity entails vanishing cutoff in Planck units \cite{Stout:2022phm} by the equivalence principle, generically Wilson coefficients measured in Planck units should diverge. Our calculation shows that, at least in weakly coupled type II sectors, when (a proxy for) the cutoff $\Lambda_\text{UV}$ diverges there exists an infinite tower of (exponentially \cite{Ooguri:2024ofs}) light states. Furthermore, translating the results of \cite{Ooguri:2024ofs} to our setting, such limits lie at infinite distance. Since we work with generic internal CFTs, this is a structural result about string perturbation theory (at least in sectors of this type) which realizes the swampland distance conjecture \cite{Ooguri:2006in} from factorization in gravity \cite{Stout:2022phm}. Let us emphasize that the necessity of a light tower at infinite distance, which here we derive from the assumption $\Lambda_\text{UV} \ll M_\text{Pl}$ motivated by information-theoretic factorization, goes in the converse direction with respect to the results obtained in \cite{Baume:2020dqd, Perlmutter:2020buo, Baume:2023msm, Ooguri:2024ofs}. Having motivated the existence of a light tower, we now take it as an assumption to investigate the precise scaling of the Wilson coefficient controlled by $I(t)$. We shall see that, under fairly generic simplifying assumptions, the behavior of geometric large-volume limits is reproduced even in (possibly) non-geometric settings.

\subsection{Light tower implies divergent Wilson coefficient}\label{sec:light_cutoff}

As anticipated in the preceding discussion, in order to study infinite-distance scalings we assume a few properties of the internal CFT:

\begin{enumerate}
    \item It has a conformal manifold with an infinite-distance limit parametrized by a marginal deformation $t$ along a geodesic. We normalize $t$ in terms of the spectral gap for convenience. We do not need fix any particular normalization of $t$ relative to the Zamolodchikov metric, although the information-theoretic considerations of \cite{Stout:2022phm} indicate that it will asymptote to $\frac{dt^2}{t^2}$ in our conventions. The explicit form of the metric will not play a role in our analysis. This dovetails with the exponential scaling in the proper distance derived in \cite{Ooguri:2024ofs}.

    \item The spectrum of its reduced torus partition function $\mathcal{Z}_{T^2}$ defined in \cref{eq:reduced_partition_function_def} contains conformal dimensions $\Delta(t)$ which are either \emph{light}, with $\Delta(t) = \Delta_* f(t)$ and $f(t) \sim \frac{1}{t}$, or \emph{heavy}, with $\Delta(t) \overset{t \gg 1}{\gg} 1$. In \cref{sec:non-factorized_limits} we assume this requirement, whereas in \cref{sec:factorized_limits} we relax it imposing this condition only on a factor in $\mathcal{Z}_{T^2}$. It would be interesting to further relax this condition to encompass more involved situations.

    \item These states are \emph{not pathological} in the following sense: firstly, the spins $j$ of the light states, which are integers by (T-)modular invariance, are all constant for some $t > T$. Were this not the case, there would be arbitrarily high-frequency step functions of $t$ in the spin spectrum. Secondly, the heavy conformal dimensions have well-behaved derivatives, in the sense that $\abs{\partial_t^k \Delta}$ grows subexponentially with $\Delta$. Functions whose derivatives grows spectacularly faster than themselves, such as (co)sine functions with (super-)exponentially growing frequency, for example feature unphysical increasingly high-frequency oscillations.
\end{enumerate}

In particular, the second assumption above captures the emergence of a light tower, a core aspect of the swampland distance conjecture \cite{Ooguri:2006in, Ooguri:2018wrx}, which in the (holographic) CFT context was reformulated and studied in \cite{Baume:2020dqd, Perlmutter:2020buo, Basile:2022sda, Baume:2023msm, Ooguri:2024ofs}. More generally, in examples internal sectors with light and heavy weights are mixed with sectors of constant weights, as \emph{e.g.} in twisted compactifications. Since the methods we employ in \cref{sec:factorized_limits} do not hinge on this subtlety, we expect our results to apply in these cases as well. Indeed, the difference between twisted and untwisted partition functions is (exponentially) suppressed, as in \cite{Scherk:1979zr, Ferrara:1987es, Kounnas:1989dk} and more recently pointed out in \cite{Abel:2024twz}. In the preceding subsection we showed that this conjecture holds if the (proxy of the) UV cutoff diverges in the limit, as it (generically) should on information-theoretic grounds \cite{Stout:2022phm}. The stronger condition on the dichotomy of light and heavy states is meant to capture the simplest kind of ``isotropic decompactification'' limit (dictated by a unique tower), and will be relaxed to factorized settings in \cref{sec:factorized_limits}. Of course, in this setting the emergent string limit is transparently singled out by $g_s \ll 1$, as we shall see below. Let us emphasize that our analysis does not rely on supersymmetry, as opposed to a great deal of investigations of the emergent string conjecture\footnote{Exceptions include \cite{Basile:2022zee} from a top-down perspective, and \cite{Basile:2023blg, Bedroya:2024ubj} from a bottom-up perspective.}. In this section we will lighten the notation, denoting the $d$-dimensional Planck scale by $M_\text{Pl} = \ell_\text{Pl}^{-1}$ and the ten-dimensional Planck scale by $M_{10} = \ell_{10}^{-1}$. 

\subsubsection{Non-factorized limits as full decompactifications}\label{sec:non-factorized_limits}

Let us begin studying the simplest settings where the (reduced) spectrum splits into light and heavy states, which ought to mimick isotropic large-volume limits without assuming geometric backgrounds. Our starting point to study the integral in \cref{eq:modular_integral_reg} is the remarkable Laplacian equation \cite{Obers:1999um, Angelantonj:2011br, Maloney:2020nni, Benjamin:2021ygh}
\begin{eqaed}\label{eq:narain_laplacian_eq}
    \left( \Delta_\tau - w_c - \Delta_\mathcal{M}\right) \Gamma_{c,c} = 0
\end{eqaed}
satisfied by Narain lattice sums, where $w_c \equiv \frac{c}{2}\left(1-\frac{c}{2}\right)$, $\Delta_\tau \equiv -y^2(\partial_x^2+\partial_y^2)$ is the Laplacian operator on the fundamental domain $\mathcal{F}$ and $\Delta_\mathcal{M}$ is the Laplacian operator on the conformal manifold. Absent any regularization issues, a convergent modular integral would yield an eigenfunction of $\Delta_\mathcal{M}$, and indeed the $SO(c,c)$ Eisenstein series briefly reviewed in \cref{sec:species_compactifications} satisfy this property. As for the general case discussed above, for $c\geq 2$ the lattice sum is not of rapid decay, but the Rankin-Selberg-Zagier procedure can be implemented subtracting the appropriate real-analytic Eisenstein series as explained above. By analogy with this well-understood setting, we would like to derive a similar equation for the general case, exploiting the assumptions about the infinite-distance limit $t \gg 1$. \\

In order to do so, we split the contributions due to light and heavy states, writing $\mathcal{Z}_{T^2}=Z_L+Z_H$. The summations over weights will be implicitly over the respective subsets. Since we have no explicit knowledge of the conformal manifold or its geometry, the aim is to exploit the assumption on light and heavy states to derive an \emph{asymptotic} differential equation resembling \cref{eq:narain_laplacian_eq}. To this end, we consider a general second-order differential operator
\begin{eqaed}
    \mathcal{D} = -A(t)\partial_t^2 - B(t)\partial_t
\end{eqaed}
acting on functions of $t$ in the $t \gg 1$ limit. Since the spins are independent of $t$, the action $(\Delta_\tau - w_c - \mathcal{D})Z_L$ of the relevant combination of operators on the $Z_L$ contains a sum over states of $e^{2\pi i j x}e^{-2\pi \Delta(t)y}$ accompanied by extra factors linear and quadratic in $y$. One straightforwardly verifies that requiring that both factors vanish in the $t \gg 1$ limit, where the conformal dimensions $\Delta(t) \sim \frac{\Delta_*}{t}$, \emph{uniquely} fixes $\mathcal{D}$ to
\begin{eqaed}
    \mathcal{D}_c \equiv - t^2\partial_t^2 - (2-c)t \partial_t \, .
\end{eqaed}
The sought-after asymptotic differential equation would thus be
\begin{eqaed}\label{eq:non-factorized_asymp_eq_integrand}
    \mathcal{D}_c \mathcal{Z}_{T^2} \sim (\Delta_\tau - w_c) \mathcal{Z}_{T^2} \, ,
\end{eqaed}
which for $c>2$ would also automatically hold for $\widetilde{\mathcal{Z}_{T^2}}$ since the above operators annihilate $E_{\frac{c}{2}}$. For $c=2$ one gets a constant source term, since $(\Delta_\tau - w_2)\widehat{E}_1 = -\frac{3}{\pi}$ \cite{Benjamin:2021ygh}. This is crucial in order to obtain the logarithmic threshold terms in the (potentially non-geometric analog of) decompactifications to eight dimensions. \\

To show that \cref{eq:non-factorized_asymp_eq_integrand} holds, we need to show that subleading terms in $Z_L$ and all contributions in $Z_H$ vanish. We begin with the former. After applying $(\Delta_\tau - w_c - \mathcal{D}_c)$, the relevant sum over the light spectrum rearranges into
\begin{eqaed}
    \sum_{j, \Delta} e^{2\pi i j x} e^{-2\pi \Delta(t) y} \left[ 2\pi y \left(c \Delta - t^2 \partial_t^2 \Delta - (2-c) t \partial_t \Delta \right) - (2\pi y)^2 \left( \Delta^2 - (t \partial_t \Delta)^2\right) \right] .
\end{eqaed}
Since for any weight in the sum $\Delta = \Delta_* f(t)$ with $f(t) \sim 1/t$, the brackets vanish termwise in the $t \gg 1$ limit. Making use of \cref{eq:abs_bound} and of the fact that the density of states of conformal weights $\rho(\Delta)$ is asymptotically power-like\cite{Afkhami-Jeddi:2020hde} due to modular invariance\footnote{More detailed expressions for spin-resolved density of states, obtained applying (a version of \cite{HeathBrown+1996+149+206}) the Rademacher circle method \cite{Rademacher:1937a, Rademacher:1937b, Rademacher:1938} can be found in \cite{Alday:2019vdr, Afkhami-Jeddi:2020hde}.}, with $\rho(\Delta) \overset{\Delta \gg 1}{\sim} \Delta^{c-1}$ (up to a prefactor), one can bound the sum in absolute value by a suitable constant prefactor times an integral of the form
\begin{eqaed}
    y^{\frac{c}{2}}\int_0^\infty d\Delta \, \Delta^{c-1} \times \Delta^k \, e^{-2\pi \Delta y} = \frac{\Gamma(c+k)}{(2\pi)^{c+k}y^{\frac{c}{2}+k}}
\end{eqaed}
for $k=1\, , \, 2$, and similarly the regularized modular integral thereof by the strip integral over $\{y \geq \frac{\sqrt{3}}{2}\}$ with measure $\frac{dy}{y^2}$. All in all, by dominated convergence the subleading contributions vanish in the limit. As for the heavy states, applying the operator to $Z_H$ once more produces a sum involving weights and derivatives thereof, which by assumption 3 can be bounded by an arbitrarily small exponential in the weight. Thus we can reuse the preceding argument since we integrate over the cut-off strip due to modular invariance. \\

As a result of these ``Tauberian'' considerations, \cref{eq:non-factorized_asymp_eq_integrand} yields the asymptotic differential equation
\begin{eqaed}
    \mathcal{D}_c I \overset{t \gg 1}{\sim} - w_c \, I
\end{eqaed}
for $c>2$, whose general solution is
\begin{eqaed}\label{eq:non-factorized_sol}
    I(t) \overset{t \gg 1}{\sim} t^{\frac{c}{2}} \left(a + \frac{b}{t}\right)
\end{eqaed}
with $a \, , \, b$ constants. The dominant scaling in the generic case is $t^{\frac{c}{2}}$, which can be suggestively rewritten as $I \sim (\sqrt{\Delta_\text{gap}})^{-c}$ in terms of the spectral gap of the reduced CFT. Since in geometric settings $\sqrt{\Delta_\text{gap}} = m_\text{gap}$ is the Kaluza-Klein scale (in string units), this scaling \emph{behaves in the same fashion as an isotropic large-volume limit!} A more cavalier way of phrasing this is that the proxy for the EFT cutoff $\Lambda_\text{UV}$ and the species scale $\Lambda_\sp$ obtained from these $R^4$ corrections indicates the emergence of geometry. We are not quite done, however: clearly $ab \neq 0$, but in principle $a=0$ is possible and the scaling would be modified in a fine-tuned subspace of solutions. To see that this does not occur, one can exploit modular invariance once again via the Rankin-Selberg-Zagier technique. The partition function for scalar states can be written as
\begin{eqaed}
    Z^0 \equiv \int_0^1 dx \, \mathcal{Z}_{T^2} = y^{\frac{c}{2}} \, F(y f(t)) + Z^0_H
\end{eqaed}
in terms of a function $F$ whose leading behavior at large $y$ is constant (precisely unity). Subtracting this behavior according to Zagier's prescription, the modified Rankin-Selberg transform $R_s[\mathcal{Z}]$ is lower bounded by the light contribution
\begin{eqaed}\label{eq:rankin-selberg_scaling}
    \int_0^\infty dy \, y^{\frac{c}{2}+s-2} \left(F(y f(t)) - 1\right) \overset{t \gg 1}{\sim} \left(\frac{t}{T}\right)^{\frac{c}{2}-1+s} R_s[Z_L|_{t=T}] \, .
\end{eqaed}
The large-$t$ scaling relative to some fixed $t=T$ arises from a change of variables in the integral $y f(t) = u f(T)$, and therefore the renormalized modular integral, given by a residue at $s=1$ of the complete Rankin-Selberg transform, scales at least like $t^{\frac{c}{2}}$, up to contributions from the heavy states. Combining \cref{eq:rankin-selberg_scaling} with \cref{eq:non-factorized_sol}, one finds the desired scaling
\begin{eqaed}
    I(t) \overset{t \gg 1}{\sim} t^{\frac{c}{2}} \propto \left(\sqrt{\Delta_\text{gap}}\right)^{-c}
\end{eqaed}
up to a constant (non-vanishing) prefactor. As for the case $c=2$, as discussed above the ``anomalous'' contribution arising from the Kronecker limit $\widehat{E}_1$ brings along an additive constant to the equation, resulting in the general solution
\begin{eqaed}
    I_{c=2}(t) \overset{t \gg 1}{\sim} a \, t + b + k \, \log t \, ,
\end{eqaed}
with $k$ fixed by the anomalous contribution. This additional term reproduces the logarithmic threshold terms in eight dimensions \cite{Green:2010wi, Angelantonj:2011br, Benjamin:2021ygh}, which may be ascribed to a logarithmic moduli dependence of the unitarizing one-loop non-analytic terms.

\subsubsection{Factorized limits as partial decompactifications}\label{sec:factorized_limits}

The above discussion appears to capture isotropic large-volume limits, and putative non-geometric generalizations thereof. In this section we relax the requirement of ``isotropy'' (the absence of conformal weights which do not vanish nor diverge as $t \to \infty$), extending our considerations to (potentially) non-geometric analogs of partial decompactifications. For the sake of simplicity we consider settings in which the reduced partition function factorizes as
\begin{eqaed}\label{eq:factorized_Z_AB}
    \mathcal{Z}_{T^2}(t) = A(t)B \, ,
\end{eqaed}
with $A$ carrying the $t$-dependence and subject to the assumptions stated at the beginning of the section. We denote by $c_1$ and $c_2$ the respective central charges, which sum to $c_1+c_2 = c = 10 - d$. Hence, the large-$y$ scalings of $A$ and $B$, dictated by the respective vacuum states, are given by $y^{\frac{c_1}{2}}$ and $y^{\frac{c_2}{2}}$ respectively. As we have anticipated in \cref{sec:light_cutoff}, the methods we apply can be extended to more general twisted settings, in which \cref{eq:factorized_Z_AB} is replaced by a linear combination of factorized terms.\\

Based on the analysis of the preceding section, one would expect that the renormalized modular integral $I_{AB}$ scale as $t^{\frac{c_1}{2}}$ up to threshold logarithms. We shall now show that this is indeed the case, except for an additional term which matches the results of \cite{Angelantonj:2011br} for partial decompactifications. To begin with, we rewrite
\begin{eqaed}\label{eq:integral_AB}
    I_{AB} \equiv \int_{\mathcal{F}}d\mu \, \widetilde{AB} & = \int_{\mathcal{F}}d\mu \left[ (\widetilde{A}+E_{\frac{c_1}{2}})(\widetilde{B}+E_{\frac{c_2}{2}})-E_{\frac{c_1+c_2}{2}}\right] \\
    & = \int_{\mathcal{F}}d\mu \left[ \widetilde{A} \widetilde{B} + \widetilde{A} E_{\frac{c_2}{2}} \right] + \text{const.}
\end{eqaed}
in order to exploit the fact that the renormalized modular integral $I_A \sim t^{\frac{c_1}{2}}$ up to a constant non-vanishing prefactor, due to the results of the preceding section. Once more, the above expression implicitly includes a Kronecker regularization for $c_2=2$. This quantity appears as the zero-mode in the harmonic expansion \cite{Benjamin:2021ygh}
\begin{eqaed}
    \widetilde{A} = \frac{3}{\pi} \, I_A(t) + \sum_{n>0} a_n(t) \, \nu_n + \int_{\text{Re}(s)=\frac{1}{2}}\!\!\!\!\!\!\!\! ds \, \alpha_s(t) \, E_s
\end{eqaed}
in terms of Maass cusp forms and real-analytic Eisenstein series. To see that the other Fourier coefficients are subleading in $t$, one can observe that for $c_1 > 2$ the asymptotic differential equation $(\mathcal{D}_{c_1} + w_{c_1})\widetilde{A} \sim \Delta_\tau \widetilde{A}$ results in
\begin{eqaed}
    (\mathcal{D}_{c_1} + w_{c_1})\int_\mathcal{F} d\mu \, \widetilde{A}E_s \sim \int_\mathcal{F} d\mu \, \widetilde{A} \Delta_\tau E_s = s(1-s) \int_\mathcal{F} d\mu \, \widetilde{A}E_s \, ,
\end{eqaed}
with $s(1-s) = \frac{1}{4} + \lambda^2$, $\lambda \in \mathbb{R}$, and analogously for the Maass cusp forms $\nu_{n>0}$ whose Laplacian eigenvalues are again in this set. As a result, the Fourier coefficients obey an asymptotic differential equation of the type
\begin{eqaed}
    \mathcal{D}_{c_1} \int_\mathcal{F} d\mu \, \widetilde{A} E_s \overset{t \gg 1}{\sim} -\left(w_c - \, \frac{1}{4} - \, \lambda^2\right) \int_\mathcal{F} d\mu \, \widetilde{A} E_s \, ,
\end{eqaed}
and analogously for the Maass cusp forms. This yields a general scaling given by linear combinations of $t^{\frac{c_1}{2}-\frac{1}{2}\pm i \lambda}$, which is indeed subleading. For $c_1=2$ the asymptotic differential equation features an additive constant, which translates to a constant particular solution, which is again subleading. \\

Inserting the harmonic expansion into the integral $I_{AB}$ of \cref{eq:integral_AB} is safe for the first term, since $\widetilde{B}$ is exponentially suppressed at large $y$. The second term needs to be handled with more care: for $c_2 \neq 2$, repeating the same Laplacian argument as above yields an asymptotic differential equation for $\int_\mathcal{F} d\mu \, \widetilde{A} E_{\frac{c_2}{2}}$ whose general solution contains $t^{\frac{c_1+c_2-2}{2}}$, which for $c_2=1$ is subleading but for $c_2 > 2$ is superleading. As anticipated in \cref{sec:stringy_rescue}, this matches the findings of \cite{Angelantonj:2011br} for partial decompactifications of a single dimension of large radius $r$ in string units, where the Rankin-Selberg transform behaves as
\begin{eqaed}
    R_s[\Gamma_{n,n}] \sim r \, R_s[\Gamma_{n-1,n-1}] & + 2\zeta^\star(2s)\zeta^\star(2s+n-2) \, r^{2s+n-2} \\
    & + 2\zeta^*(2s-1)\zeta^*(2s-n+1) \, r^{n-2s}
\end{eqaed}
up to exponentially suppressed corrections in $r$, where $\zeta^\star(s) \equiv \frac{\Gamma\left(\frac{s}{2}\right)}{\pi^{\frac{s}{2}}} \zeta(s)$. The residue at $s=1$ gives a term proportional to $r = t^{\frac{c_1}{2}}$ and another term proportional to $r^{n-2} = t^{\frac{c_1+c_2-2}{2}}$. The prefactor must be carefully computed for each $n \geq 4$ (the relevant case since $c_2 > 2$ and $c_1 > 0$), since in involves a product of Euler gamma and Riemann zeta functions. These superleading contributions correspond to operators whose gap-scale suppression appears ``additively'' with respect to the Planck scale in the $d$-dimensional EFT. To see this, it is more convenient to switch to a geometric language, writing $c_1 = n$, $c_2 = k$ for internal dimensions and $t^{\frac{n}{2}} = \mathcal{V} = \left(\frac{M_\text{Pl}}{M_s}\right)^{d-2}$ for the decompactifying volume in string units. Inserting these expressions with the superleading scaling into the full Wilson coefficient yields
\begin{eqaed}\label{eq:superleading_wilson_coefficient}
    \alpha \sim \left(\frac{M_\text{Pl}}{\Lambda_\sp}\right)^{8-d} \mathcal{V}^{1 + \frac{k-2}{n}} = \left(\frac{M_\text{Pl}}{\Lambda_\sp}\right)^{(8-d) \frac{n+d-2}{n}} = \left(\frac{m_\text{gap}}{M_\text{Pl}}\right)^{d-8} ,
\end{eqaed}
thus leading to a gap-scale suppression of the ``additive'' type, namely without any appearance of $M_\text{Pl}$. This analysis, which can be extended to all operators of dimension larger than $d+n$, further corroborates the arguments presented in \cref{sec:stringy_rescue}, extending them to more general internal CFTs. \\

Finally, for $c_2=2$, once more there is an additive contribution, now proportional to $-\frac{3}{\pi} \, I_A(t)$, which brings along the expected logarithmic threshold behavior $\log t$. In this case the general solution is a linear combination of $t^{\frac{c_1}{2}}$, $t^{\frac{c_1}{2}-1}$ (subleading with respect to the zero-mode $I_A I_B$) and a threshold scaling $t^{\frac{c_1}{2}} \log t$ with a fixed prefactor, which perfectly reproduces the $\mathcal{V} \log \mathcal{V}$ geometric behavior in decompactification limits to eight dimensions \cite{Green:2010wi, Angelantonj:2011br, Benjamin:2021ygh}.

\section{Conclusions}\label{sec:conclusions}

The basic physical question driving this work is whether string theory does require extra dimensions in the classical sense. More precisely, whether classical geometry is always accessible in its perturbative moduli space of CFT backgrounds. Its web of stringent consistency conditions implies something \emph{a priori} more general, which brings along a number of deep ideas about what the correct notion of ``geometry'' in quantum gravity may be. However, attempting to connect string theory to low-energy phenomenology, one would like to identify ubiquitous infrared features of the string landscape. If perturbative transitions to geometric extra dimensions are always present, perhaps the corners of the landscape explored hitherto are larger than one may have previously imagined, and predictions from these corners may generalize. \\

The detailed interplay of string dualities leading to the dichotomy of infinite-distance limits encapsulated by the emergent string conjecture of \cite{Lee:2018urn, Lee:2019wij, Lee:2019xtm} suggests, among other things, that the above picture is correct, since in perturbative string theory one can neatly single out the limits in which a weakly coupled string dominates the physics. The rest then ought to be described by Kaluza-Klein degrees of freedom in some duality frame. This claim is already non-trivial in geometric settings, due to (T-)dualities, and our investigation indicates that it applies more broadly. In particular, combined with the analysis in \cref{app:geometric_vacua} and in \cite{Ooguri:2024ofs}, it would also apply in the presence of a continuous family of quasi-de Sitter vacua accumulating at zero dark energy, along the lines of \cite{Montero:2022prj}. \\

The bottom-up analyses of \cite{Basile:2023blg, Bedroya:2024ubj} dovetail nicely with our top-down results in this respect. Furthermore, as discussed in the note added to \cref{sec:introduction}, the novel results of \cite{Ooguri:2024ofs}, interpreted from a top-down worldsheet perspective rather than the intended bottom-up holographic one, further corroborate this picture, showing that limits in conformal manifolds where light states emerge lie at infinite distance and exhibit exponential scalings (consistently with the information-theoretic arguments of \cite{Stout:2022phm}\footnote{In order to reach this conclusion, it seems that at least in $d\geq 4$ it is important that the infrared spacetime physics be Einstein gravity \cite{Basile:2022sda}. The holographic interpretation of \cite{Ooguri:2024ofs} suggests that in $d=3$ this may not be the case.}) and the limiting theory contains a decompactified (free) sigma model sector. \\

Our results on the asymptotics of UV-sensitive scales along the limit constitute another step toward clarifying whether geometry emerges in string theory in the sense of perturbative connectedness. Namely, the scale we have investigated vanishes in Planck units if and only if a light tower arises due to modular invariance, and it does so in a geometric fashion. More precisely, our proof of the former implication assumes the presence of an internal sector whose conformal weights are either (uniformly) light or heavy as the spectral gap vanishes, and whose behavior is not wildly oscillatory. Our approach can be extended to other higher-derivative operators which may exhibit different scalings, other sectors of string theory (such as heterotic), and to more complicated infinite-distance limits, in which multiple internal CFT factors can scale differently. Collectively, these bottom-up and top-down results provide encouraging indications that, after all, extra dimensions might be ubiquitous in the landscape, and that the remarkable dichotomy of infinite-distance limits in gravity does persist.

\appendix
\section{Local heat kernel expansion}\label{app:heat}

In this appendix we briefly review the local, diagonal heat kernel expansion \cite{Vassilevich_2003}. Let $\mathcal{M}$ be an $n$-dimensional smooth, compact Riemannian manifold with boundary $\partial \mathcal{M}$ and let $V\rightarrow \mathcal{M}$ be a generic vector bundle. We say that a differential operator $D:V\rightarrow V$ is of \textit{Laplace-type} if it can be locally represented as a second-order differential operator acting on (sections of) $V$ with appropriate boundary conditions on $\partial\mathcal{M}$. In particular, it can be shown that there exists a unique endomorphism $E$ of $V$ such that $D$ has the following local representation:
\begin{eqaed}
    D = -\left(I\,g^{\mu \nu}\nabla_{\mu} \nabla_{\nu} + E\right) \, ,
\end{eqaed}
where $g$ is the Riemannian metric on $\mathcal{M}$, $\nabla$ the covariant derivative with respect to the connection on $V$ and the Levi-Civita connection, and $I$ is the identity on $V$. With this, and given an auxiliary function $f \in C^{\infty}(\mathcal{M})$, one can build a trace-class one by exponentiating, meaning that the quantity
\begin{eqaed}\label{eq:heatkerneltracedef}
    K_{V}(t,f) \equiv \mathrm{Tr}_{V}\left[ f \, e^{-tD}\right] ,
\end{eqaed}
where the (super)trace is meant to be taken over the space of square-integrable functions $L^2(V)$, is well-defined. Introducing the kernel $K(t,x,y)$ of the heat equation constructed from $D$, we can also write it as
\begin{eqaed}
     K_{V}(t,f) = \int_{\mathcal{M}}d^nx\sqrt{g} \, \mathrm{tr}_V K(t,x,x) f(x) \, ,
\end{eqaed}
relating the trace over the vector bundle to the functional trace on $\mathcal{M}$ and the trace $\mathrm{tr}_V$ over the vector space given by the canonical fiber of $V$. For this reason, \eqref{eq:heatkerneltracedef} has been given the name of \textit{heat kernel trace}. While impractical to compute explicitly outside of simple cases, the heat kernel trace admits an asymptotic expansion for $t\rightarrow 0^+$ \cite{seeley1966singular,seeley1969resolvent,Vassilevich_2003} which we can write as
\begin{eqaed}
     K_{V}(t,f) \sim \frac{1}{(4\pi t)^{\frac{n}{2}}}\sum_{k \geq 0} \left[ a_k^\mathcal{M}(f) + a_k^{\partial\mathcal{M}}(f) \right] t^{\frac{k}{2}} \, ,
\end{eqaed}
where the $a_k(f)$ are called Seeley-DeWitt coefficients or heat kernel coefficients. These coefficients usually depend on the integral of various local curvature invariants of the bundle $V$ and of the tangent bundle $T\mathcal{M}$. The contribution from the boundary $a_k^{\partial \mathcal{M}}$ also depends on the boundary conditions imposed, and we will not report or use it. Instead, we report here for the reader's convenience the first few coefficients in the simple case where $D = \Delta$, the Laplace-Beltrami operator on $\mathcal{M}$, $\partial\mathcal{M} = \emptyset$ and the bundle $V$ is trivial, so that we can only focus on curvature invariants constructed from the Riemann tensor. The coefficients read \cite{Vassilevich_2003}
\begin{eqaed}
    &a_0^\mathcal{M}(f) = \int_\mathcal{M}d^n x \, \sqrt{g} \, f(x) \, ,\\
    &a_2^\mathcal{M}(f) = \int_\mathcal{M}d^n x \, \sqrt{g} \, R \, f(x) \, , \\
    &a_4^\mathcal{M}(f) = \int_\mathcal{M}d^n x \,\sqrt{g}\, f(x) \left( 12 \Delta R + 5R^2-2R^{\mu \nu}R_{\mu\nu} + 2R^{\mu\nu\rho\sigma}R_{\mu\nu\rho\sigma}\right) \, ,\\
    &a_{2k+1}^{\mathcal{M}}(f)=0 \, .
\end{eqaed}
Notice that in the case where $f=1$, the first term reduces to the overall volume $V$ of the internal manifold.
\\

It is also worth reviewing the role of the heat kernel expansion in computing one-loop effective actions, a useful tool employed in \cref{sec:endangered}. Assume that the Laplace-type operator $D$ defined above (or possibly an extension on non-compact manifolds \cite{Vassilevich_2003}) appears in the quadratic term for some fields (which are thus sections of the appropriate bundles) in the classical action $S$, and denote the fields collectively as $\phi$. The (Euclidean) path integral for the quantum theory reads
\begin{eqaed}
    Z = \int \mathcal{D}\phi \exp \left( -S[\phi] \right),
\end{eqaed}
whereas formally expanding around a classical solution $\phi = \phi_\text{cl} + \chi $ up to second order, we have:
\begin{eqaed}
    Z = e^{-S[\phi_\text{cl}]} \int\mathcal{D}\chi \exp \left( -\chi \cdot \frac12\frac{\delta^2 S}{\delta \chi^2} \cdot \chi \right) + \mathcal{O}(\chi^3)\, ,
\end{eqaed}
where the dot denotes a functional product. Identifying $D = \frac{\delta^2 S}{\delta \chi^2}$, introducing the effective action $W =-\log Z $ and performing the Gaussian integral, we have that
\begin{eqaed}
    W = S[\phi_\text{cl}] + S_{\text{1-loop}}[\phi_\text{cl}] + \mathcal{O}(\chi^3) \, , \quad  S_{\text{1-loop}}[\phi_\text{cl}] = \frac{1}{2}\log \det D \, .
\end{eqaed}
Using the identity
\begin{eqaed}
 \log \frac{a}{b} = -\int_0^{\infty}\frac{ds}{s} \left(e^{-sa}-e^{-sb}\right)
\end{eqaed}
on every eigenvalue of $D$, we find the final expression in terms of the heat kernel trace\footnote{The operator $D$, being Laplace-type, is self-adjoint on compact manifolds (with appropriate boundary conditions) and affords a spectral decomposition in terms of its eigenfunctions, which can be used to build the trace.}:
\begin{eqaed}
    S_{\text{1-loop}} = -\frac{1}{2}\int_{\Lambda^{-2}}^\infty \frac{dt}{t} \, K_V(t,D) \, ,
\end{eqaed}
where we have introduced a UV regulator $\Lambda$ to keep parametric control over divergences. As a last remark, it is interesting to evaluate the case where the manifold (and bundles) factorize according to $\mathcal{M} = M \times X$, and so does the operator $D = D_M + D_X$, in terms of spacetime $M$ and a compact internal manifold $X$. The functional spaces and the traces factorize, and one can express the one-loop effective action as
\begin{eqaed}
    S_{\text{1-loop}} = -\,\frac{1}{2}\int_{\Lambda^{-2}}^\infty \frac{dt}{t}K_X(t,D_X)K_M(t,D_M) \, ,
\end{eqaed}
where the heat kernel expansion for $K_M$ yields a covariant form of the one-loop contribution to the effective action of the reduced theory on $M$ in terms of operators defined by the heat kernel coefficients $a_k^{M,\partial M}(\mathcal{R})$, where $\mathcal{R}$ collectively denotes (local) curvature invariants of the various bundles. As a result, one obtains
\begin{eqaed}
    S_{\text{1-loop}} \sim  -\frac{1}{2(4\pi)^{\frac{d}{2}}}\int_{\Lambda^{-2}}^\infty \frac{dt}{t^{1+\frac{d}{2}}}K_X(t,D_X)\sum_{k \geq 0} \left[ a_k^M(\mathcal{R}) + a_k^{\partial M}(\mathcal{R}) \right] t^{\frac{k}{2}} \, .
\end{eqaed}

\section{Leading-order equivalence of species counting regulators}\label{app:tao}

In this appendix we briefly present an argument by Tao \cite{Tao11} extended from compactly supported regulators to a suitable class which includes the exponential regulator discussed in \cref{sec:species_compactifications}. An extension to Schwartzian functions was presented in \cite{Padilla:2024mkm}\footnote{We are grateful to A. Padilla for pointing this out to us.}. The aim is to show the leading-order equivalence between sharp and smooth regulators, as well as the universality of the latter for arbitrary choices, in the species counting of \cref{eq:speciesdef}. Up to a suitable redefinition of the smooth regulator $\eta$, we first recast the two sums as
\begin{eqaed}
        &N_\sp = \sum_{n=1}^{N_\sp} 1, \\
        & N_\sp^{\text{smooth}}= \sum_{n=1} \eta\left(\frac{n}{N_{\sp}}\right) . 
\end{eqaed}
Then, we estimate $N_\sp^{\text{smooth}}$ explicitly. Consider the generic Taylor expansion
\begin{eqaed}
    f(n + 1) = f(n) + f'(n) + \mathcal{O}(\norm{f}_{C^2}),
\end{eqaed}
where we have bounded Lagrange's remainder by using the $C^2$-norm $\norm{f}_{C^2} = \mathrm{sup}_{x\in \mathbb{R}}|f''(x)|$. Then, employing the same kind of Taylor expansion we can also compute
\begin{eqaed}
    \int_n^{n+1}dx f(x) = f(n) + \frac{1}{2} \, f'(n) + \mathcal{O}(\norm{f}_{C^2}) \, ,
\end{eqaed}
while eliminating $f'(n)$ with the above identities we arrive at
\begin{eqaed}
    \int_n^{n+1}dx f(x) = \frac12 f(n) + \frac{1}{2} \, f(n+1) + \mathcal{O}(\norm{f}_{C^2}) \, ,
\end{eqaed}
and summing up to $N$ we find
\begin{eqaed}
    \int_0^{N}dx f(x)= \frac{1}{2} \, f(0) + f(1) + \dots + f(N-1) + \frac{1}{2} \, f(N) + \mathcal{O}(N\norm{f}_{C^2}).
\end{eqaed}
Applying this to $f(x) = \eta({x}/{N_{\sp}})$ and identifying $N=N_\sp$, using the fact that due to the chain rule $\mathcal{O}(\norm{\eta}_{C^2}) = \mathcal{O}(N_{sp}^{-2})$ we obtain
\begin{eqaed}\label{eq:etaint}
    N_\sp\int_0^{1} dy \, \eta\left( y\right) = \frac{1}{2} + \sum_{n=1}^{N_\sp-1}   \eta\left( \frac{n}{N_\sp}\right) +\frac{1}{2} \, \eta(1)+ \mathcal{O}\!\left(\frac{1}{N_\sp}\right),
\end{eqaed}
where we conventionally chose $\eta(0) = 1$ and we performed a change of variables in the integral. We can use the same trick to write the integral of $\eta$ in the range $(N_\sp,\infty)$, now using the fact that the function and all of its derivatives vanish rapidly after $N_\sp$. This yields
\begin{eqaed}
    \int_{N_\sp}^\infty dx \, \eta \left(\frac{x}{N_\sp} \right) = \frac{1}{2} \, \eta(1) + \sum_{n=N_\sp + 1}^{\infty}\eta\left(\frac{n}{N_\sp} \right) + \mathcal{O}\!\left(\frac{1}{N_\sp}\right),
\end{eqaed}
where we imposed a suitable condition implicitly defining the class of rapidly vanishing functions we are interested in, namely
\begin{eqaed}\label{eq:rapidvanish}
    \sum_{n=N_\sp}^\infty\mathcal{O}\left(\norm{\eta}_{C^2_n}\right) = \mathcal{O}\!\left(\frac{1}{N_\sp}\right),
\end{eqaed}
where the norm is taken for every interval $[n,n+1]$, that is $||\eta||_{C^2_n} \equiv \mathrm{sup}_{[n,n+1]}\abs{\eta''}$ . Using this, and rearranging terms in \eqref{eq:etaint}, we finally get to
\begin{eqaed}
    N_\sp^{\text{smooth}} = -\frac{1}{2} + C N_\sp + \mathcal{O}\!\left( \frac{1}{N_\sp}\right),
\end{eqaed}
where $C$ is the integral of $\eta$ over all $\mathbb{R^+}$. This result shows that in the limit where $N_\sp \gg 1$, the smooth regulator coincides with the sharp one when extracting the leading-order answer, up to an irrelevant multiplicative constant.

\section{Worldsheet distances in geometric vacua}\label{app:geometric_vacua}

In \cref{sec:species_compactifications} we have 
discussed general geometric limits in the context of EFT, showing that no qualitative difference with respect to the toroidal case arises in the species and cutoff scales for average internal curvatures much lower than the higher-dimensional Planck scale. In this appendix we connect this picture with the worldsheet approach of \cref{sec:species_worldsheet}, discussing distances from the point of view of the latter incorporating curvature deformations and showing agreement with the EFT approach in the geometric regime. We shall see that the generalized field-space metric in the sense of \cite{Lust:2019zwm}\footnote{See also \cite{Basile:2022zee, Basile:2022sda, Basile:2023rvm, Li:2023gtt, Palti:2024voy} for related approaches in this context.} reduces to the DeWitt expression \cite{PhysRev.160.1113, gilmedrano1992riemannian} for parametrically small curvatures, while allowing for systematic computation of stringy effects. The relevant information metric \cite{oconnor, Lassig:1989tc, dolan1, Dolan:1995zq, Stout:2021ubb, Stout:2022phm} for the string worldsheet is the Zamolodchikov metric \cite{Zamolodchikov:1986gt, Cvetic:1989ii, Friedan:2012hi}\footnote{An alternative proposal, along the lines of \cite{Douglas:2010ic}, was explored in \cite{Bachas:2013nxa}.}, since conformal invariance on a flat worldsheet follows from Weyl and diffeomorphism invariance of the Polyakov path integral construction. The following expressions are valid at tree level in string perturbation theory, assuming the backgrounds are consistent so that one may gauge away the worldsheet metric and the dilaton-Ricci scalar term with it. \\

For a family of target-space metrics $G_{\mu \nu}(X \,|\, \lambda)$, the Zamolodchikov metric coefficients are given by connected correlators of the form
\begin{eqaed}\label{eq:quinfo_metric_family_spacetimes}
   g_{ij}(\lambda) = \int_{\mathcal{W}} \star \langle (\partial_{\lambda^i}G_{\mu \nu}(X) \, \partial X^\mu \cdot \partial X^\nu) (\partial_{\lambda^j}G_{\rho \sigma}(X) \, \partial X^\rho \cdot \partial X^\sigma)  \rangle_{\text{c}} \, .
\end{eqaed}
As opposed to the well-understood metric on Narain conformal manifolds, this more general expression can in principle account for curvature corrections, \emph{e.g.} in (critical) Schwarzschild-Tangherlini metrics parametrized by the Schwarzschild radius $r_{\text{s}}$, at least at leading order in $\alpha'$. The worldsheet approach allows systematic computation of $\alpha'$ corrections to the EFT distance of \cite{Lust:2019zwm}, at least when the background exists to all orders (\emph{e.g.} for Calabi-Yau sigma models \cite{Nemeschansky:1986yx, Jardine:2018sft}). However, while a number of non-trivial marginal deformations are integrable \cite{Sfetsos:2013wia, Hollowood:2014rla}, there seems to be an obstruction to exact solvability of marginal curvature deformations, since they can be recast as a factorized free-boson algebra\footnote{We thank E. Kiritsis for bringing this point to our attention.}. \\

Varying the NS-NS $B$-field leads to a straightforward generalization of \cref{eq:quinfo_metric_family_spacetimes},
\begin{eqaed}\label{eq:quinfo_metric_family_NS-NS}
   \partial_{\lambda^i}G_{\mu \nu}(X) \, \partial X^\mu \cdot \partial X^\nu \quad \longrightarrow \quad ((\partial_{\lambda^i}G_{\mu \nu}(X)\delta^{\alpha \beta} + i \, \partial_{\lambda^i}B_{\mu \nu}(X)\epsilon^{\alpha \beta}) \, \partial_\alpha X^\mu \partial_\beta X^\nu)
\end{eqaed}
on a Euclidean worldsheet. More generally, since the parameters appear in the couplings of the Lagrangian (density) $\mathcal{L}_{\lambda}$,
\begin{eqaed}\label{eq:more_general_metric}
   g_{ij}(\lambda) = \int_{\mathcal{W}} \star \langle \partial_{\lambda^i}\mathcal{L}_{\lambda} \, \partial_{\lambda^j}\mathcal{L}_{\lambda} \rangle_{\text{c}} \, .
\end{eqaed}
These expressions can be evaluated in worldsheet perturbation theory (as an asymptotic series in the curvature(s) of the background) employing Riemann normal coordinates. Namely, one redefines the worldsheet fields according to the geodesic expansion around a \emph{constant} background $\overline{X}^\mu(\sigma) = x_0^\mu$, and the leading interactions in the fluctuation fields $Y^a$ (with indices in the tangent space obtained via the background vielbein, as in~\cite{Callan:1989nz}, so that the kinetic terms are canonical) considerably simplify to the schematic structure
\begin{eqaed}\label{eq:schematic_worldsheet_interactions}
   (\partial Y)^2 + \text{Riem} \cdot Y Y \partial Y \partial Y + H \cdot \epsilon \, Y \partial Y \partial Y \, .
\end{eqaed}
The curvatures of the background depend on $\lambda$, and the leading metric, easily computed via Wick contractions, is a linear combination of the schematic contractions $\partial_{\lambda^i} \text{Riem} \cdot \partial_{\lambda^j} \text{Riem}$ and $\partial_{\lambda^i} H \cdot \partial_{\lambda^j} H$ (the mixed terms vanish). \\

The above discussion is valid when the topology of target space is trivial and the marginal perturbations modify the curvature. To cover settings such as toroidal compactifications, where changing radii does not affect the curvatures, one has to take into account the different structure of the free CFT for the geodesic fluctuations $Y$. For simple cases, such as Narain theories, one can treat moduli explicitly, \emph{e.g.} for $S^1_R$ writing kinetic terms as $R^2 \partial X \cdot \overline{\partial} X$ (and similarly for a general complex structure $\tau$ of a two-torus or higher-dimensional Narain deformations). More general cases may be harder to treat with this approach, and instead using canonical kinetic terms with fields on non-trivial target spaces may be preferable. In this case, moduli do not affect the derivatives of the classical action, but they do affect the free energy contribution to the quantum information metric, since the free correlators depend on moduli $\{t \}$ such as internal radii. In order to take this into account, we split curvature deformations and ``flat'' moduli, letting uppercase indices $I, J$ denote the curvature-related parameters $\lambda$ and the flat moduli $t$ collectively, and denoting with $F^a(\lambda) \int \mathcal{O}_a(Y)$ the perturbations containing the $\lambda$. In this approach, when the deformations are linear in the couplings, one can express the metric in terms of the Hessian matrix of the (negative) free energy density. Then one can separate the contributions of the ``flat'' moduli $\{ t \}$ treating the curvature deformations $\lambda$ as perturbations. Then, from conformal perturbation theory the general formula for the metric coefficients $g_{IJ}(\lambda, t)$ to the first nontrivial order is
\begin{eqaed}\label{eq:master_metric_moduli}
    g_{IJ}(\lambda, t) & = g^{\text{free}}_{IJ}(t) +    \frac{1}{2} \, \partial_I F^a \partial_J F^b \int \langle \mathcal{O}_a \mathcal{O}_b \rangle_t + \frac{1}{2} \, F^a \, F^b \int \partial_I \partial_J \langle \mathcal{O}_a \mathcal{O}_b \rangle_t \\
    & + \frac{1}{2} \,  F^a \partial_I F^b \partial_J \int \langle \mathcal{O}_a \mathcal{O}_b \rangle_t + \frac{1}{2} \,  F^a \partial_J F^b \partial_I \int \langle \mathcal{O}_a \mathcal{O}_b \rangle_t \, ,
\end{eqaed}
where the $t$ subscript denotes (connected) moduli-dependent free correlators. In the case at hand, $F^a$ denotes either $\text{Riem}$ or $H$ (up to proportionality constants) and the corresponding operators are schematically $Y Y \partial Y \overline{\partial} Y$ and $Y \partial Y \overline{\partial} Y$ in Euclidean worldsheet complex coordinates. In general, one of the upshots is that distances in the flat moduli space can be modified by curvature, due to the last term in the first row of the above expression.

\subsection{Leading-order curvature expansion}\label{sec:low_curvature_worldsheet}

Concretely, for the nonlinear sigma model at leading order in the curvatures (and their spacetime derivatives) we can use complex coordinates on the Wick-rotated worldsheet and expand the action
\begin{eqaed}\label{eq:NLSM_complex_coords}
   S = \frac{1}{2\pi \alpha'}\int d^2\sigma \left(G_{\mu \nu}(X) + B_{\mu \nu}(X) \right) \partial X^\mu \overline{\partial}X^\nu 
\end{eqaed}
in fluctuations $Y^a$ with tangent-space indices to obtain
\begin{eqaed}\label{eq:NLSM_expanded_complex_coords}
   \frac{1}{2\pi \alpha'} \int d^2z \left(\partial Y^a \overline{\partial}Y^a + \frac{1}{3} R_{abcd} Y^b Y^c \partial Y^a \overline{\partial}Y^d + \frac{1}{3} H_{abc} Y^a \partial Y^b \overline{\partial}Y^c\right) \, .
\end{eqaed}
The resulting metric requires the two-point correlators of the (implicitly normal-ordered) composite operators
\begin{eqaed}\label{eq:required_correlators}
   \mathcal{O}_R^{abcd} & \equiv Y^b Y^c \partial Y^a \overline{\partial} Y^d \, , \\
   \mathcal{O}_H^{abc} & \equiv Y^a \partial Y^b \overline{\partial} Y^c \, ,
\end{eqaed}
and we now restrict to trivial target space topology, so that we can use the standard CFT correlators. The metric components, up to an irrelevant prefactor\footnote{This may include the difference between intensive and extensive metrics (see \cite{Stout:2021ubb, Stout:2022phm}) and/or a proper normalization relevant for the swampland distance conjecture, for instance involving the Planck scale or the string coupling.}, then read
\begin{eqaed}\label{eq:more_explicit_metric}
   g_{ij}(\lambda) & = \partial_{\lambda^i}R_{abcd} \partial_{\lambda^j}R_{pqrs} \int \!d\abs{z} \, \abs{z} \,\langle \mathcal{O}_R^{abcd} \mathcal{O}_R^{pqrs} \rangle_\text{c}^{\text{free}} \\
   & + \partial_{\lambda^i}H_{abc} \partial_{\lambda^j}H_{pqr} \int \!d\abs{z} \, \abs{z} \,\langle \mathcal{O}_H^{abc} \mathcal{O}_H^{pqr} \rangle_\text{c}^{\text{free}} \, ,
\end{eqaed}
where we used rotational invariance of the correlators and the integrals are implicitly regulated with a UV length scale $a$ (say with a lattice-type regularization). This result can be compared with the proposed expressions for distances in the space of metrics, perhaps upon performing a similar geodesic/curvature expansion. The latter term in \cref{eq:more_explicit_metric} is the simplest, and it gives
\begin{eqaed}\label{eq:dHdH_correlator}
   - 2 \left( \frac{\alpha'}{2} \right)^3 \, \partial_{\lambda^i}H_{abc} \partial_{\lambda^j}H^{abc} \int \!d\abs{z} \, \frac{\log(\frac{\abs{z}}{a}) + 1}{\abs{z}^3} \, .
\end{eqaed}
The lattice regularization amounts to the replacement $1/\abs{z} \to \abs{z}/(\abs{z}^2 + a^2)$, so that after changing variables the integral evaluates to
\begin{eqaed}\label{eq:reg_dHdH_integral}
   \frac{1}{a^2} \int_0^\infty \left( 1 + \log\left(u + \frac{1}{u}\right) \right) \frac{u^3}{(1+u^2)^3} du = \frac{1}{2a^2} \, .
\end{eqaed}
As expected, it follows from dimensional analysis that the result is proportional to $a^{-2}$, so that indeed the intensive metric is finite. The second term thus contributes $- \left( \frac{\alpha'}{2} \right)^3 \partial_{\lambda^i}H_{abc} \partial_{\lambda^j}H^{abc}$. The result is negative because the corresponding operator, with our conventions, is anti-Hermitian, since it arises from a topological (here meaning metric-independent) coupling. The correct $B$-field rescaled by a factor of $i$ makes the resulting action Hermitian, and the information metric positive. \\

The Riemann correlator $\left(\frac{2}{\alpha'}\right)^4\abs{z}^4\langle \mathcal{O}_R^{abcd} \mathcal{O}_R^{pqrs} \rangle_\text{c}^{\text{free}}$, after some tedious algebra, evaluates to\footnote{We thank J. Freigang for cross-checking these calculations in the early stages of development of this work.}
\begin{eqaed}\label{eq:OROR_correlator}
    & \left(\log (\abs{z}/a)^2\right)^2 \left(\delta^{ap} \delta^{bq} \delta^{cr} \delta^{ds} + \delta^{ap} \delta^{br} \delta^{cq} \delta^{ds}\right) \\
    - & \log (\abs{z}/a)^2 \left(\delta^{aq} \delta^{bp} \delta^{cr} \delta^{ds} + \delta^{ar} \delta^{bp} \delta^{cq} \delta^{ds} + \delta^{ap} \delta^{bs} \delta^{cq} \delta^{dr} + \delta^{ap} \delta^{bs} \delta^{cr} \delta^{dq} \right) \\
    - & \log (\abs{z}/a)^2 \left(\delta^{aq} \delta^{br} \delta^{cp} \delta^{ds} + \delta^{ar} \delta^{bq} \delta^{cp} \delta^{ds} + \delta^{ap} \delta^{bq} \delta^{cs} \delta^{dr} + \delta^{ap} \delta^{br} \delta^{cs} \delta^{dq} \right) \\
    + \; & \delta^{aq} \delta^{bp} \delta^{cs} \delta^{dr} + \delta^{ar} \delta^{bp} \delta^{cs} \delta^{dq}  + \delta^{aq} \delta^{bs} \delta^{cp} \delta^{dr} + \delta^{ar} \delta^{bs} \delta^{cp} \delta^{dq} \, .
\end{eqaed}
Using the symmetries of the Riemann tensor of the (torsionless) Levi-Civita connection, after contraction one arrives at
\begin{eqaed}\label{eq:dRdR_correlator}
   2 \left( \frac{\alpha'}{2} \right)^4 \left(\partial_{\lambda^i}R_{abcd} \partial_{\lambda^j}R^{abcd} + \partial_{\lambda^i}R_{abcd} \partial_{\lambda^j}R^{acbd} \right)\int \!d\abs{z} \, \frac{2\log^2(\frac{\abs{z}}{a}) + 2\log(\frac{\abs{z}}{a}) + 1}{\abs{z}^3} \, ,
\end{eqaed}
and the lattice-regularized integral evaluates to
\begin{eqaed}\label{eq:reg_dRdR_integral}
   \frac{1}{a^2} \int_0^\infty \left( 1 + 2\log\left(u + \frac{1}{u}\right) + 2\log^2\left(u + \frac{1}{u}\right)\right) \frac{u^3}{(1+u^2)^3} du = \frac{3/4 + 2J}{a^2} \equiv \frac{C}{a^2} \, ,
\end{eqaed}
where $J \equiv \int_0^\infty \log^2\left(u + \frac{1}{u}\right) \frac{u^3}{(1+u^2)^3} \, du \approx 0.294383$. Therefore, to leading order in the curvatures $\text{Riem}$ and $H$, their spacetime derivatives\footnote{There is a term $\nabla H$ in the background-field expansion multiplying a quartic operator in $Y$ that carries the same power of $\alpha'$ as the Riemann term. Since we assume slowly-varying curvatures $\text{Riem}$ and $H$, we can suppress this term relative to the ones in \cref{eq:final_bosstring_metric_leading}.} and the string coupling, and for trivial spacetime topology, up to an irrelevant prefactor the metric reads
\begin{eqaed}\label{eq:final_bosstring_metric_leading}
    g_{ij}(\lambda) = \alpha' C \left(\partial_{\lambda^i}R_{abcd} \partial_{\lambda^j}R^{abcd} + \partial_{\lambda^i}R_{abcd} \partial_{\lambda^j}R^{acbd} \right) + \partial_{\lambda^i}H_{abc} \partial_{\lambda^j}H^{abc}
\end{eqaed}
with $C \equiv \frac{3}{4} + 2J \approx 1.33877$. \\

Focusing on spacetime distance variations, at first glance it would seem that the above expression does not quite match the DeWitt one: expanding the latter about a point in Riemann normal coordinates, one obtains schematically
\begin{eqaed}\label{eq:de_witt_comparison}
    & \partial_{\lambda^i}R_{abcd} \partial_{\lambda^j}{{R^a}_{pq}}^d \, \frac{1}{\text{Vol}}\int d^Dx \, \, x^b x^c x^p x^q \\
    & \propto \partial_{\lambda^i}R_{abcd} \partial_{\lambda^j}R^{abcd} + \partial_{\lambda^i}R_{abcd} \partial_{\lambda^j}R^{acbd} + \partial_{\lambda^i}R_{ad} \partial_{\lambda^j}R^{ad} \, .
\end{eqaed}
However, for \emph{on-shell} variations, at this order in $\alpha'$ and $g_s$ the Ricci tensor vanishes along the conformal manifold, and therefore the Zamolodchikov metric does coincide with the DeWitt one. To remove the spurious dependence on the background position due to the perturbative expansion, one can average over the \emph{background} position, at least for compact spaces. The full answer should not change, while the leading-order answer loses the dependence on the arbitrary choice of background. The resulting expression is then
\begin{eqaed}\label{eq:averaged_metric}
    \overline{g}_{ij}(\lambda) = \frac{1}{\text{Vol}} \int \left( \alpha' C \left(\partial_{\lambda^i}R_{abcd} \partial_{\lambda^j}R^{abcd} + \partial_{\lambda^i}R_{abcd} \partial_{\lambda^j}R^{acbd} \right) + \partial_{\lambda^i}H_{abc} \partial_{\lambda^j}H^{abc} \right) \, ,
\end{eqaed}
where the volume factor may also depend on $\lambda$. \\

The above discussion of worldsheet distances in geometric backgrounds dovetails with the heat-kernel estimates of \cref{sec:species_compactifications}, showing that there is no qualitative disagreement to the story due to the introduction of small curvatures. Therefore, in order to seek significant deviations from the pattern of decompactification and emergent string limits, it is natural to turn to more abstract CFT methods, as discussed in \cref{sec:species_worldsheet}.

\section*{Acknowledgements}

We are grateful to C. Angelantonj, R. Blumenhagen, A. Castellano, G. Di Ubaldo, A. Herr\'{a}ez, D. L\"{u}st, E. Kiritsis, C. Kneissl, A. Paraskevopoulou and N. Risso for insightful discussions, and S. Raucci for feedback on the manuscript. \\
This work is supported through the grants CEX2020-001007-S and PID2021-123017NB-I00, funded by MCIN/AEI/10.13039/501100011033 and by ERDF A way of making Europe. C.A. is supported by the grant PID2021-123017NB-I00.

\bibliographystyle{ytphys}
\baselineskip=.95\baselineskip
\bibliography{ref}

\end{document}